\documentclass[aps,prx,twocolumn,superscriptaddress]{revtex4-2}

\usepackage{amsmath}
\usepackage{siunitx}
\usepackage[inline]{enumitem}

\usepackage{xcolor}  % For corrections
\usepackage{graphicx}
\usepackage{amsfonts}
\usepackage{booktabs}
\usepackage{xcolor}
\usepackage{mathtools}
\usepackage{upgreek, fixmath,amssymb, mathtools}

\newtheorem{result}{Result}

\DeclarePairedDelimiter\bra{\langle}{\rvert}
\DeclarePairedDelimiter\ket{\lvert}{\rangle}
\DeclarePairedDelimiter\norm{\Vert}{\Vert}
\DeclarePairedDelimiter\abs{\vert}{\vert}
\NewDocumentCommand{\tens}{e{_^}}{%
  \mathbin{\mathop{\otimes}\displaylimits
    \IfValueT{#1}{_{#1}}
    \IfValueT{#2}{^{#2}}
  }%
}

\graphicspath{{Images/}}

\begin{document}

%\preprint{}

\title{Boundary scattering tomography of the Bose Hubbard model on general graphs}


\author{Abhi Saxena}
\thanks{co-first authors}
\affiliation{Department of Electrical \& Computer Engineering, University of Washington; Seattle, Washington, 98195, USA}
\author{Erfan Abbasgholinejad}
\thanks{co-first authors}
\affiliation{Department of Electrical \& Computer Engineering, University of Washington; Seattle, Washington, 98195, USA}
\author{Arka Majumdar}
\affiliation{Department of Electrical \& Computer Engineering, University of Washington; Seattle, Washington, 98195, USA}
\affiliation{Department of Physics, University of Washington; Seattle, Washington, 98195, USA}
\author{Rahul Trivedi}
\email[Corresponding author: ]{rtriv@uw.edu}
\affiliation{Department of Electrical \& Computer Engineering, University of Washington; Seattle, Washington, 98195, USA}
%\thanks{}


\date{\today}

\begin{abstract}

Correlated quantum many-body phenomena in lattice models have been identified as a set of physically interesting problems that cannot be solved classically. Analog quantum simulators, in photonics and microwave superconducting circuits, have emerged as near-term platforms to address these problems. An important ingredient in practical quantum simulation experiments is the tomography of the implemented Hamiltonians --- while this can easily be performed if we have individual measurement access to each qubit in the simulator, this could be challenging to implement in many hardware platforms. In this paper, we present a scheme for tomography of quantum simulators which can be described by a Bose-Hubbard Hamiltonian while having measurement access to only some sites on the boundary of the lattice. We present an algorithm that uses the experimentally routine transmission and two-photon correlation functions, measured at the boundary, to extract the Hamiltonian parameters at the standard quantum limit. Furthermore, by building on quantum enhanced spectroscopy protocols that, we show that with the additional ability to switch on and off the on-site repulsion in the simulator, we can sense the Hamiltonian parameters beyond the standard quantum limit.

\end{abstract}

%\keywords{}

\maketitle


\section{Introduction} 

\begin{figure*}[!htbp]
	\centering
	\includegraphics[width=1\linewidth]{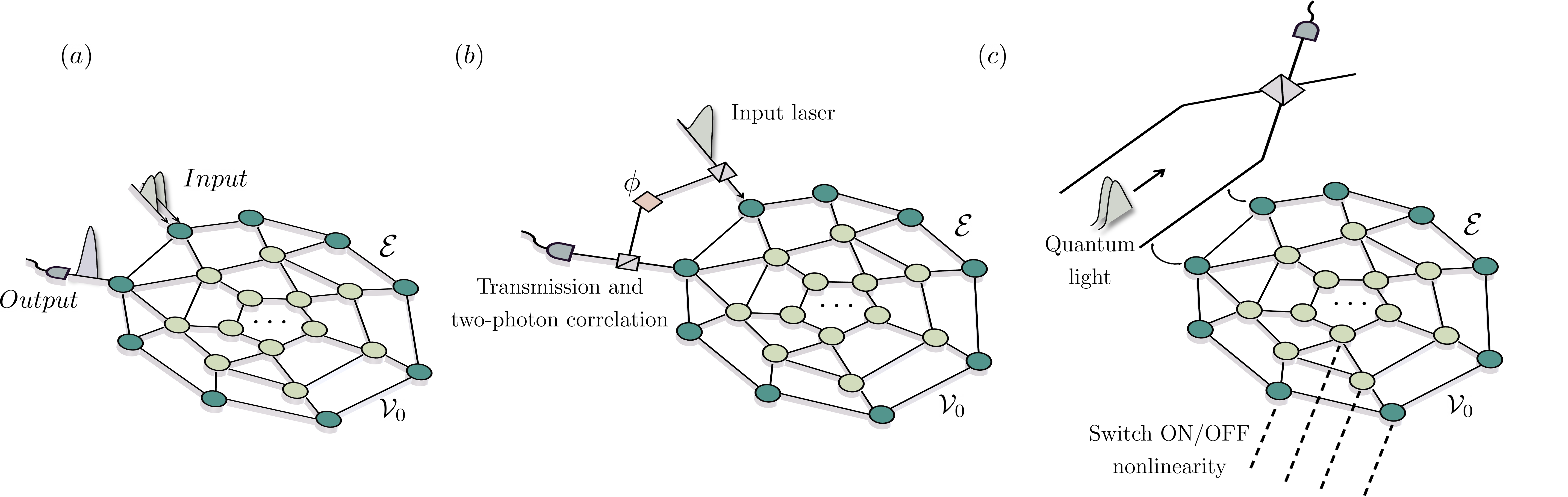}
	\caption{\label{fig:fig1001} (a) Schematic depicting the model of the photonic graph ($\mathcal{V}, \mathcal{E}$). The set $\mathcal{V}_0 \subseteq \mathcal{V}$ denotes the set of vertices forming the outermost boundary which are accessible to measurement. (b) Measurements used for Result 1 --- one of the boundary sites is excited with a weak laser and the emission from another boundary site is homodyned with the laser and then detected with either a transmission measurement, or a two-photon correlation measurement. (c) Measurements used for Result 2 --- we excite boundary sites coupled to an arm of an interferometer with a designed Fock state or NOON state, and measure the scattered photon. For this protocol, we additionally assume the ability to switch on and off the nonlinearity in the model.}
\end{figure*}

Correlated quantum many-body effects are often hard to simulate on classical computers and have been a subject of considerable recent interest as potential problems to probe with quantum simulators. There has been considerable progress in engineering interacting bosonic modes with onsite non-linearity in superconducting systems or photonics \cite{carusotto2020photonic, hartmann2016quantum, anderson2016engineering, owens2022chiral, schine2016synthetic, hafezi2013imaging}, and this has made these some of the most promising analogue quantum simulation platforms for studying many-body physics. An important ingredient in the goal of using quantum simulators are protocols that allow accurate tomography of the implemented Hamiltonians \cite{hangleiter2021precise} with only modest measurement and initial state preparation apparatus. If the experimental system allows measurement and state preparation access to arbitrary sites in the lattice, direct tomography can be performed to characterize the implemented Hamiltonian \cite{ma2017hamiltonian, haah2023optimal}. However, in many experimentally relevant settings \cite{saxena2022realizing, mittal2019photonic, kim2021quantum}, implementing measurement access to a node in the interior of the lattice is significantly more challenging than the nodes on the outermost boundary can be accessed for accurate measurements. This experimental requirement raise an important theoretical question --- is it possible to perform tomography of the implemented Hamiltonian with access restricted to only the boundaries of the lattice? 

This question has been considered in previous works, which have proposed tomography algorithms with restricted measurement access to the boundary or a subpart of the lattice. In the setting of one-dimensional chains  as well as for more general graphs, tomography algorithms for single-particle coupling coefficients in several excitation number preserving models with only boundary measurement access have been proposed \cite{burgarth2009coupling, di2009hamiltonian, burgarth2009indirect, maruyama2012application, burgarth2011indirect} --- these algorithms, however, assume that all the coupling parameters in the Hamiltonian are real and positive, and thus are not applicable to many topologically non-trivial models \cite{stormer1999fractional, hafezi2011robust, hafezi2013imaging, owens2022chiral}. Subsequent work has been done to perform tomography of spin lattices with real-valued couplings and restricted number of probes through measurement of just the system time-traces, both for closed \cite{zhang2014quantum} and open quantum systems \cite{zhang2015identification}. Further, the identifiability of the Hamiltonian using the protocol has also been studied to determine its applicability to various models \cite{sone2017hamiltonian, wang2020quantum}. Additionally, work has been done to improve the efficiency of the tomography algorithms using Bell states \cite{burgarth2012quantum}, identify one-dimensional Hamiltonians through Zeeman markers \cite{burgarth2017evolution} and perform entanglement tomography of many-body Hamiltonians \cite{kokail2021entanglement}.  

 However, there still remain several experimentally relevant open questions --- \emph{first}, none of these protocols would not be suitable for identification of topological many body Hamiltonians which are typically characterized by complex single-particle coupling strengths between modes to realize non-zero flux in closed loops \cite{hafezi2011robust, hafezi2013imaging, owens2022chiral}. \emph{Second}, quantum simulators simulating a `classically hard' model necessarily have non-gaussian terms (such as two-particle repulsion) in their Hamiltonian --- it remains unclear how strengths of these non-gaussian terms can be measured with just boundary measurement access. \emph{Finally}, all the existing tomography algorithms achieve a precision in the reconstructed Hamiltonian parameters at the standard quantum limit (SQL) \cite{giovannetti2006quantum} --- it is unclear if a quantum advantage over SQL can be achieved in sensing all the Hamiltonian parameters, as can be achieved for sensing single phase or frequency parameters in the usual setting of quantum spectroscopy.
 
 In this work, we address all of these questions in the setting where the quantum simulator can be modelled with a Bose Hubbard Hamiltonian --- this model applies to several existing quantum simulation platforms such as superconducting qubit systems, photonic quantum simulators as well as cold-atoms in optical lattices \cite{gemelke2009situ, hartmann2016quantum, ma2019dissipatively}. By building on previous results by Burgarth et al.~\cite{burgarth2009indirect, burgarth2011indirect}, which is already applicable to the Bose Hubbard model with real-valued coupling strengths and no on-site nonlinearities, we show that both the \emph{complex valued} single-particle couplings and onsite potentials can be reconstructed from transmission measurements. Furthermore, we show that even the onsite anharmonicity can be reconstructed from two-photon correlation function measurements from the boundary sites and under coherent-state excitation of the boundary sites. Our results are applicable to Bose-Hubbard models on graphs that satisfy the same graph infection condition that was identified and used by Burgarth et al in Refs.~\cite{burgarth2009indirect, burgarth2011indirect}. Furthermore, we then investigate the possibility of going beyond the standard quantum limit using non-classical excitations. By building on the protocols for quantum enhanced spectroscopy \cite{pezze2018quantum, bollinger1996optimal, holland1993interferometric, sanders1989quantum}, we show that if we are able to toggle the nonlinearity in the Hamiltonian, then using multi-photon Fock state scattering from the boundaries of the lattice allows us to quantum-enhance the precision in the measured Hamiltonian parameters.
 
 % To the best of our knowledge, our work is the first to solve this problem for general topologically non-trivial bosonic lattices with anharmonic lossy resonators and complex couplings. Moreover, as our tomography algorithm depends on steady-state measurements in the frequency domain, it avoids the need: (i) to measure time-resolved traces, which can be challenging for strongly coupled systems, (ii) to utilize initial state preparation, or (iii) to perform quick qubit rotations. Several of the schemes mentioned earlier in the last paragraph suffer from one or more of these potential pitfalls. While our method unlike some quantum Hamiltonian tomography algorithms \cite{ma2017hamiltonian, hangleiter2021precise} is not applicable for arbitrary photonic lattice geometries it allows precise identification of most common topological Hamiltonian models like the Hofstadter lattices \cite{hafezi2011robust, hafezi2013imaging, owens2022chiral, owens2018quarter}, strained graphene Hamiltonians \cite{youssefi2022topological}, quantized quadrupole phase lattices \cite{mittal2019photonic} and SSH Hamiltonians \cite{saxena2022photonic, mittal2019photonic} without requiring experimental access to all the nodes of the lattice.  

\section{Setup and Summary of results}\label{Model}
\emph{Setup}: We consider $N$ bosonic modes arranged on a graph ($\mathcal{V}, \mathcal{E}$), where $\mathcal{V}$ is the vertex set signifying the bosonic modes and $\mathcal{E}$ are edges which signify the linear couplings between the bosonic modes [Fig. \ref{fig:fig1001} (a)]. The Hamiltonian modelling this system is given by: 
\begin{align*}
% \label{hamiltonian}
	H =\sum_{v \in \mathcal{V}}  \bigg(\omega_v a_v^\dag a_v + \frac{\chi_v}{2} a_v^{\dag 2} a_v^2\bigg) + \sum_{u,v \in \mathcal{E}} J_{v,u}a_v^\dag a_u  
\end{align*}
where the parameters $J_{v,u} \in \mathbb{C}$  (the coupling strengths which also can be assumed to satisfy $ J_{v,u}= J^*_{u,v}$), $\omega_v \in \mathbb{R} $ (resonant frequency), $\chi_v \in \mathbb{R} $ (the onsite anharmonicity) need to be measured. Furthermore, in an experimentally realistic setting, each bosonic mode will also suffer from particle decay, which will impact the photon transport spectrum of the system. Including this loss, it has been previously shown that \cite{xu2015input, trivedi2018few, trivedi2019photon} the few-photon transport spectrum is determined by the effective non-Hermitian Hamiltonian
\begin{align}\label{eq:effective_hamiltonian_introduction}
\displaystyle
H_\text{eff} & = H -\frac{i}{2}\sum_{v \in \mathcal{V}} \kappa_v a_v^\dag a_v \notag \\ & = \sum_{v \in \mathcal{V}}  \bigg( \mu_v a_v^\dag a_v + \frac{\chi_v}{2} a_v^{\dag 2} a_v^2 \bigg) + \sum_{u,v \in \mathcal{E}} J_{v,u}a_v^\dag a_u
\end{align}
where $\mu_v = \omega_v - i\kappa_v / 2$. In this case, our goal would be to reconstruct the loss rates $\kappa_v$ in addition to the parameters of the lossless Hamiltonian. Throughout this paper, we will assume that we have boundary access to this lattice of bosonic modes. This is depicted schematically in Fig.~\ref{fig:fig1001} --- $\mathcal{V}_0$ is the set of vertices that correspond to the bosonic modes on the boundary of the lattice, and we will assume the ability to excite and measure the bosonic modes at these vertices. This assumption is motivated from existing experimental limitations in most quantum hardware platforms, where setting up measurement access to every qubit is difficult \cite{saxena2022realizing, mittal2019photonic, kim2021quantum}. 

\begin{figure*}[!htbp]
    \centering
    \includegraphics[width=0.9\linewidth]{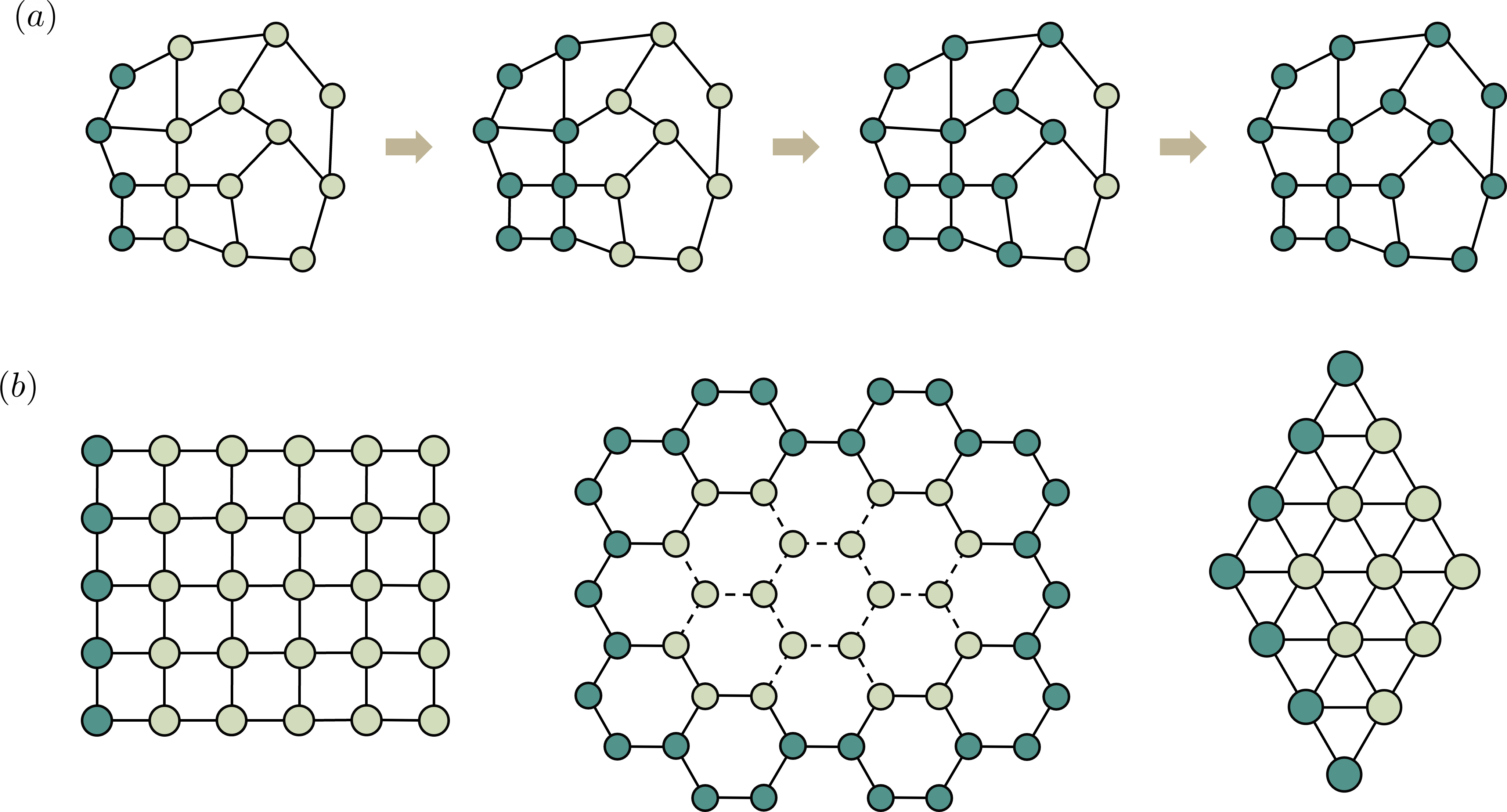}
    \caption{(a) Schematic illustration of the graph infection condition --- the vertices with measurement access, shown in a dark green color, are assumed to be initially infected. These vertices then infect their unique nearest neighbours --- the graphs that we consider in this paper should be such that, proceeding this way, all the vertices can be infected. (c) Examples of graphs, as well as choice of ``boundary vertices" (i.e.~initially infected vertices) that satisfy the graph infection condition.}
    \label{fig:examples_assumption_1}
\end{figure*}

\emph{Summary of results.} We will consider several different kinds of boundary measurements, ranging from ones that are routinely performed in near-term experiments (Result 1, Fig.~\ref{fig:fig1001}(b)) to those that require more careful design of the excitation and detection setup (Result 2, Fig.~\ref{fig:fig1001}(c)). The first case that we consider is where can excite the boundary modes with a continuous wave coherent state (e.g.~a laser), and we have access to homodyne single-photon transmission (which allows us to reconstruct the parameters $J_{v,u}, \mu_v$) and two-photon correlation measurements (which allows us to reconstruct the parameters $\chi_v$) [Fig.~1(b)]. For both of these measurements, an input port, through which a laser excitation is applied, is coupled to a boundary site $v \in \mathcal{V}_0$ and an output port is coupled to another boundary site $u \in \mathcal{V}_0$. The input port is then excited with a continuous-wave coherent state and, furthermore, the emission from the output port is interfered with the input laser field after passing it through a controllable phase shifter before being received at a photodetector.

First, we perform a transmission measurement --- here, we excite the input port with a continuous-wave coherent state at a single frequency $\omega$, and measure the intensity at the output port after homodyning --- as we detail in appendix \ref{apxB}, it follows from photon scattering theory  \cite{xu2015input,trivedi2019photon} that a measurement of this intensity as a function of the frequency $\omega$ as well as the homodyne phase $\varphi$ allows us to extract the (complex) energies $E_\alpha^{(1)}$ of the effective Hamiltonian in Eq.~\ref{eq:effective_hamiltonian_introduction} within the single excitation subspace, as well as the parameters $M_{\alpha, v, u}^{(1)}$
\begin{align}
    M_{\alpha, v, u}^{(1)} = \bra{G} a_v \ket{r_\alpha^{(1)}} \bra{l_\alpha^{(1)}}a_u^\dagger \ket{G} \text{ for }v, u \in \mathcal{V}_0,
\end{align}
where $\ket{l_\alpha^{(1)}}$ and $\ket{r_\alpha^{(1)}}$ are the left and right single-excitation eigenstates of $H_\text{eff}^{(1)}$. The parameters $M_{\alpha, v, u}^{(1)}$, for $v, u \in \mathcal{V}_0$, contain information about the overlap of the single-excitation eigenstates of $H_\text{eff}$ with an excitation at the boundary vertices. We remark that since we use a homodyne measurement, we are able to extract both the magnitude \emph{and} phase of $M_{\alpha, v, u}^{(1)}$ --- this will be key in allowing us to reconstruct potentially complex coupling coefficients $J_{v, u}$ in the model.

Next, we perform a two-particle correlation measurement. For this measurement, we use a two-tone continuous-wave coherent state with frequencies $\omega_1, \omega_2$ through the input port and measure the two-particle correlation function at the output, after homodyning, as a function of arrival times $\tau_1$ and $\tau_2$ of the two photons. As we detail in appendix~\ref{apxB}, by varying the incident frequencies $\omega_1, \omega_2$, the homodyned phase $\varphi$ and recording the correlation function as a function of $\tau_1, \tau_2$, we can extract the complex energies $E_\alpha^{(2)}$ of the effective Hamiltonian in Eq.~\ref{eq:effective_hamiltonian_introduction} within the two-excitation subspace, as well as the parameters $M_{\alpha, v, u}^{(2)}$
\begin{align}
    M_{\vec{\alpha},v, u}^{(2)} =\langle G| a_v  |r_{\alpha_1}^{(1)}\rangle \langle l_{\alpha_1}^{(1)}| a_v  |r_{\alpha_2}^{(2)}\rangle  \langle l_{\alpha_2}^{(2)}| a_u^\dag |r_{\alpha_3}^{(1)}\rangle \langle l_{\alpha_3}^{(1)}|  a_u^\dag |G \rangle ,
\end{align}
where $\vec{\alpha} = (\alpha_1, \alpha_2, \alpha_3)$ and $|{l_\alpha^{(2)}}\rangle, |{r_\alpha^{(2)}}\rangle$ are the two-excitation left and right eigenvectors of the effective Hamiltonian. Again, we emphasize that we are able to measure both magnitude and phase of $M_{\vec{\alpha}, v, u}^{(2)}$ since we use a homodyne measurement.

In section \ref{sec:tomography_with_tran_g2}, we build on Refs. \cite{burgarth2009coupling, di2009hamiltonian, burgarth2009indirect, maruyama2012application, burgarth2011indirect} to provide an algorithm to reconstruct both the single-particle parameters $J_{v, u}$ and $\mu_v$ and provide a new algorithm to reconstruct the two-particle parameters $\chi_v$. Our algorithm works under the ``graph infection" assumption introduced in Ref.~\cite{burgarth2011indirect}, which we briefly recap below and is schematically depicted in Fig.~\ref{fig:examples_assumption_1}(a). Suppose that we start with the vertices in $\mathcal{V}_0$ as being labelled as ``infected", then we proceed to label another vertex $v \in \mathcal{V}\setminus \mathcal{V}_0$ as infected if it is the \emph{unique uninfected nearest neighbour} of an infected vertex. The algorithm that we propose for sensing the Hamiltonian parameters will succeed if all the vertices in $\mathcal{V}$ can be infected starting from the vertices in $\mathcal{V}_0$. It can be explicitly verified that several uniform lattices that are routinely encountered in quantum simulation problems, such as square, honeycomb and triangular lattices, satisfy the graph infection condition often with only a partial boundary [Fig.~\ref{fig:examples_assumption_1}(b)] However, graphs with higher density of edges between the interior and exterior vertices might not satisfy this infection condition, and would thus not be amenable to our tomographic algorithm. An example of one such graph is shown is the honeycomb lattice with nearest and next-to-nearest neighbor coupling, which appears in the Haldane model. We expect that for such graphs, boundary measurements along will not be sufficient for determining the Hamiltonian parameters uniquely.
% \begin{figure}[!htbp]
%     \centering
%     \includegraphics[width = 0.65\linewidth]{nFig3.png}
%     \caption{The honeycomb lattice with additional next-to-nearest neighbor couplings, which is often used to define the Haldane model. This is an example of a lattice where assumption 1 does not apply, and our tomography algorithm fails to reconstruct the single-particle parameters.}
%     \label{fig:examples_violation_assumption_1}
% \end{figure}
\noindent Our first result can now be stated as follows:
\begin{result}
Consider the Bose-Hubbard model on a lattice that can be infected starting from the boundary vertices $\mathcal{V}_0$, then the parameters $J_{v, u}, \mu_u$ and $\chi_u$ can be reconstructed uniquely from the energy eigenvalues $E_\alpha^{(1)}, E_\alpha^{(2)}$, and $M_{\alpha, v, u}^{(1)}, M^{(2)}_{\vec{\alpha}, v, u}$ with $v, u \in \mathcal{V}_0$, which can be inferred from homodyned transmission and two-particle correlation functions measured with only boundary access.
\end{result}
\noindent For the single particle coupling parameters $J_{v, u}$, we go significantly beyond the algorithm in Refs.~\cite{burgarth2009coupling, di2009hamiltonian, burgarth2009indirect, maruyama2012application, burgarth2011indirect} and additionally consider lattices where the phase of the coupling coefficients $J_{v, u}$ is important, while Ref.~\cite{burgarth2011indirect} assumed that the coupling coefficients are real and positive. In particular, and as explained in section \ref{sec:tomography_with_tran_g2}, the key step that allows us to handle complex value couplings is an identification of a loops of edges between one uninfected and remaining infected nodes, that allow us to fix the gauge in which the Hamiltonian parameters can be uniquely constructed. Additionally, we also provide an algorithm for the reconstruction of $\chi_v$ from two-particle correlation functions, which is an experimentally relevant setting that has not been previously explored. 

We also numerically study the stability of our reconstruction algorithm i.e.~to what precision should the measurements at the boundary vertices be performed so as to allow for a certain target precision in the reconstructed parameters. The precision required at the boundary measurement would then set the number of repetitions of the measurement that needs to be performed, or alternatively, the number of exciting photons that need to be used --- in particular, to perform a boundary measurement of the transmission or the correlation function to a precision $\delta$, we would need $O(\delta^{-2})$ photons or measurement trials. For the tomography algorithm to be efficient and scalable, it is desirable that required measurement precision scales at most poly($N$) with the number of sites. Through our numerical studies, we find that for models with delocalized single and two-particle eigenstates, the required measurement precision to determine $J_{v, u}$ and $\mu_v$ only scales polynomially with the number of sites $N$, while it scales exponentially with $N$ if the target is to determine $\chi_v$ to a target precision. This potentially limits our algorithm for reconstructing $\chi_v$ to small system sizes --- however, this can be remedied in platforms where it is possible to switch the non-linearity $\chi_v$ on each site on and off, and we display a simple algorithm that utilizes this additional control to stably find $\chi_v$. Section \ref{sec:tomography_with_tran_g2} provides a detailed description of this algorithm, together with numerical studies of its stability for some paradigmatic examples.

Next, we consider the question of a possible quantum enhancement of the precision with which the parameters of the Hamiltonian can be measured. The tomography algorithm described thus far, for a fixed number of sites $N$, only obtains a precision constrained by the standard quantum limit i.e.~in order to determine the coefficients in the Hamiltonian to a precision $\varepsilon$  would require $O(
\varepsilon^{-2})$ photons --- this is simply due to the fact that the photons are used independently while performing the measurement. We then investigate the question of improving the dependence of the number of photons on the target precision $\varepsilon$ beyond the standard quantum limit by using all the photons in parallel. Here, as depicted in Fig.~\ref{fig:fig1001}(c), we assume the ability to switch on and off the on-site two-particle repulsion term in the Hamiltonian and adapt methods used in quantum-enhanced spectroscopy \cite{pezze2018quantum, bollinger1996optimal, holland1993interferometric, sanders1989quantum} to show the following result.
\begin{result}
    Consider the Bose-Hubbard model on a lattice that satisfies assumption 1 and assume the ability to switch on and off the nonlinear term in the model. Then, for a fixed lattice size $N$, the parameters $J_{v, u}$ and $\mu_u$ can be reconstructed to a precision $\varepsilon$ with $O(\varepsilon^{-1})$ photons and only using boundary measurements. Furthermore,  $\chi_u$ can be reconstructed to a precision of $\varepsilon$ with $O(\varepsilon^{-2/3})$ photons and only using boundary measurements.
\end{result}
Section \ref{sec:quantum_enhancement} details the reconstruction protocol. We point out that while the result here builds on well-known quantum-enhanced spectroscopy protocols, such as twin-Fock state spectroscopy \cite{ bollinger1996optimal, holland1993interferometric} and NOON state spectroscopy \cite{sanders1989quantum}, these protocols usually sense a single and easily accessible system parameter (e.g.~a single phase shift, or a frequency). Our key contribution is to utilize and extend these techniques to enhance the precision of all the Hamiltonian parameters while only being able to excite and measure the lattice through the boundary vertices.

We also point out that while we attain a Heisenberg scaling (i.e.~$O(\varepsilon^{-1})$) in the number of photons required for reconstruction of the single particle parameters $J_{v, u}$ and $\mu_u$, we obtain a \emph{super Heisenberg} scaling (i.e.~slower than $O(\varepsilon^{-1}$)) in the number of photons required for the reconstruction of $\chi_v$. Such a super-Heisenberg scaling is not surprising or unphysical, and has been previously obtained for sensing problems with a many-body probe with all-to-all interactions \cite{boixo2008quantum1, boixo2008quantum2}, as well as for the problem of sensing non-linearity in a single anharmonic oscillator \cite{xie2022quantum}. As analyzed in Ref.~\cite{xie2022quantum}, this super-Heisenberg scaling arises from the fact that if a single anharmonic oscillator with $H = \chi a^{\dag 2}a^2 / 2$ is initialized in the $P$ photon state $\ket{P}\propto a^{\dag P}\ket{0}$, then after time $t$, this state acquires a phase $\propto \chi P^2 t$ that scales \emph{quadratically} with $P$. Thus, if we are able to measure this phase to a constant precision, then to obtain a precision $\varepsilon$ in $\chi$, we simply need to choose $P \sim \varepsilon^{-1/2}$. The problem of measuring $\chi_v$ in the Bose-Hubbard model, however, is more involved since the photons at any one oscillator can also interact with other oscillators, and consequently we don't accumulate a simple phase on time-evolution that can be measured later. However, as detailed in Section \ref{sec:quantum_enhancement}, this unwanted interaction can be suppressed for by choosing a sufficiently short evolution time, while still retaining a super-Heisenberg scaling of the required number of photons with the target precision in $\chi_v$.

% and in the next stage, we reconstruct the non-linear terms of the Hamiltonian ($\chi$'s) using two photon measurements.\\

\section{Tomography with single and two-particle measurements}\label{sec:tomography_with_tran_g2}

\begin{figure*}[!htbp]
	\centering
	\includegraphics[width=1\linewidth]{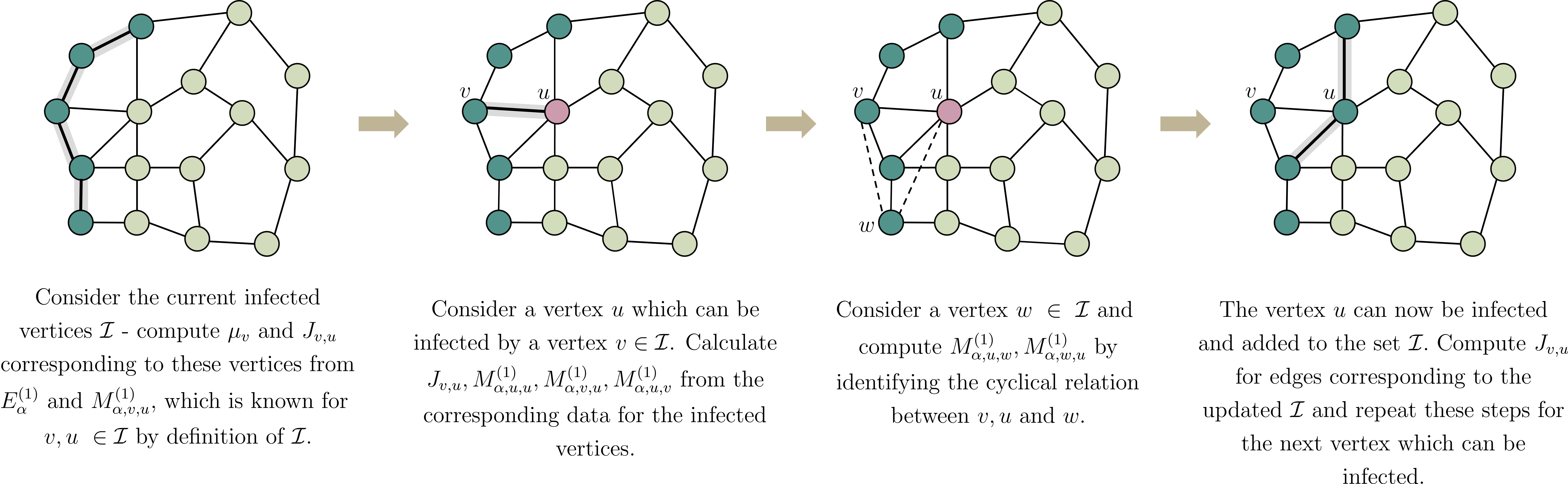}
	\caption{\label{fig:steps} Schematic depicting the tomography algorithm in the single excitation space. The dark-green vertices are the currently infected vertices, while the red vertex is the vertex being infected.}
\end{figure*}

In this section, we provide details of the tomography algorithm corresponding to result 1. We perform the tomography in two steps --- first, we consider reconstruction of only the single-particle parameters $J_{v, u}$ and $\mu_v$ from the measurements of the single-excitation eigen-energies $E_\alpha^{(1)}$ and $M_{\alpha, v, u}^{(1)}$. Having determined these parameters, we then reconstruct the two-particle parameters $\chi_v$. Finally, we will provide a numerical study of the stability of the tomography algorithm.
\subsection{Reconstructing the single-particle parameters $J_{v, u},\ \mu_v$}
In the first stage of the scheme, we use $M^{(1)}_{\alpha, v,u} \ \forall \ v,u \in \mathcal{V}_0$ and eigen-energies $E^{(1)}_\alpha$ extracted from boundary transmission measurements to reconstruct the single-particle parameters $J_{v, u}$ and $\mu_v$. As schematically shown in Fig.~\ref{fig:steps} and similar to Ref.~\cite{burgarth2009coupling}, the idea behind the reconstruction procedure is to keep track of a set of infected vertices $\mathcal{I}$ which initially is $\mathcal{V}_0$, compute the parameters $\mu_v$ and $J_{v, u}$ for $v, u \in \mathcal{I}$ and sequentially add uninfected vertices to this set till the full lattice has been covered. At any stage of the reconstruction procedure, we will define the set of infected vertices $\mathcal{I}$ to be such that $M_{\alpha, v, u}^{(1)}$ is known for all $v, u \in \mathcal{I}$ such that the edge $(v, u)$ exists. For ease of notation, we define by $\mathcal{E}_I$ the set of edges in the Hamiltonian graph that involve vertices in $\mathcal{I}$ i.e. $\mathcal{E}_I = \{(v, u) \in \mathcal{E} | v, u \in \mathcal{I}\}$. At the start of the reconstruction procedure, by assumption we have that $\mathcal{V}_0 = \mathcal{I}$. With this data together with the eigenenergies $E_\alpha^{(1)}$, we can then determine $\mu_v \in \mathcal{I}$ and $J_{v, u}$ for $(v, u) \in \mathcal{E}_I$. To This can be determined straightforwardly from the eigenvalue equation corresponding to $H_\text{eff}$ within the single excitation subspace \cite{ma2017hamiltonian}. More concretely, since $H_\text{eff}|r_\alpha^{(1)}\rangle = E_\alpha^{(1)}|r_\alpha^{(1)}\rangle$ and $\langle l_\alpha^{(1)}| H_\text{eff} = E_\alpha^{(1)} \langle l_\alpha^{(1)}|$, we have that
\begin{subequations}\label{eq:single_particle_eigval}
\begin{align}
    &E_\alpha^{(1)}\langle G | a_v |{r_\alpha^{(1)}}\rangle = \mu_v \langle G | a_v |r_\alpha^{(1)}\rangle + \sum_{v' \in \mathcal{N}_v} J_{v, v'} \langle G | a_{v'} |r_\alpha^{(1)}\rangle, \\
    &E_\alpha^{(1)}\langle l_\alpha^{(1)} | a_v^\dag \ket{G} = \mu_v \langle l_\alpha^{(1)} |a_v^\dag \ket{G} + \sum_{v'\in \mathcal{N}_v}J_{v', v}  \langle l_\alpha^{(1)}| a_{v'}^\dag \ket{G},
\end{align}
\end{subequations}
where $\mathcal{N}_v$ is the deleted neighbourhood of the vertex $v$ i.e.~it is the set of vertices other than $v$ that share an edge with $v$.
Using the completeness of $\ket{l_\alpha^{(1)}}, \ket{r_\alpha^{(1)}}$, $\sum_{\alpha} |l_\alpha^{(1)}\rangle \langle r_\alpha^{(1)}| = I^{(1)}$ (where $I^{(1)}$ is the identity within the single excitation subspace) we obtain from this equation that for every $v \in \mathcal{I}$ and $(v, u) \in \mathcal{E}_I$,
\begin{align}\label{J_uv}
\mu_v =  \sum_{\alpha} E_\alpha^{(1)}M^{(1)}_{\alpha, v, v} \ \text{and } J_{v, u} = \sum_{\alpha} E_\alpha^{(1)} M_{\alpha, v, u}^{(1)}.
\end{align}
Next, consider a vertex $u \in \mathcal{V} \setminus \mathcal{I}$ that is uninfected but can be infected by the vertices in $\mathcal{I}$. By definition, this means that $u$ is a unique uninfected nearest neighbour of a vertex $v \in \mathcal{I}$. First, we consider the coupling $J_{v, u}$ --- we can assume this coupling to be real and positive since this choice amounts to fixing the local gauge of the bosonic mode at site $u$, which has not been fixed as of yet. Now, we can rewrite Eqs.~\ref{eq:single_particle_eigval}(a) and (b) as
\begin{subequations}\label{eq:rewrite_main_eigvals}
    \begin{align}
        &J_{v, u} \bra{G}a_u \ket{r_\alpha^{(1)}} = \big(E_\alpha^{(1)} - \mu_v\big) \bra{G} a_v \ket{r_\alpha^{(1)}} -\nonumber\\ &\qquad \qquad \sum_{v' \in \mathcal{N}_v\setminus\{u\}} J_{v, v'}\langle G | a_{v'}\ket{r_\alpha^{(1)}} \ \text{and } \\
        &J_{v, u} \bra{l_\alpha^{(1)}}a_u^\dagger\ket{G} = \big(E_\alpha^{(1)} - \mu_v\big) \bra{l_\alpha^{(1)}} a_v^\dagger \ket{G} -\nonumber\\ &\qquad \qquad \sum_{v' \in \mathcal{N}_v\setminus\{u\}} J_{v, v'}^*\langle l_\alpha^{(1)}| a_{v'}^\dagger \ket{{G}},
    \end{align}
\end{subequations}
where we have used that $J_{v, u}$ is real and $J_{v, v'} = J^*_{v', v}$. Multiplying these equations and summing over $\alpha$, we obtain that
\begin{align}\label{J_v'v}
	J_{v,u}^2 = \sum_\alpha (E^{(1)}_\alpha -\mu_v)^2 M^{(1)}_{\alpha,v,v} -  \sum_{v' \in \mathcal{N}_v \setminus \{u\}}|J_{v,v'}|^2.
\end{align}
Note that $J_{v, u}$ is positive, and that $J_{v, v'}$ for all $v' \in \mathcal{N}_v \setminus \{u\}$ are known. This is so because $u$ is the only uninfected neighbour of $v$, consequently the set vertices in $\mathcal{N}_v \setminus \{u\}$ are already in $\mathcal{I}$. Therefore, $J_{v, u}$ can be determined from Eq.~\ref{J_v'v}. Furthermore, once $J_{v, u}$ have been determined, we can again use Eqs.~\ref{eq:rewrite_main_eigvals}(a) and (b) to determine $M^{(1)}_{\alpha, u, u}$, $M^{(1)}_{\alpha, v, u}$ and $M^{(1)}_{\alpha, u, v}$. In particular, $M^{(1)}_{\alpha, u, u}$ can be determined by multiplying Eqs.~\ref{eq:rewrite_main_eigvals}(a) and (b):
\begin{subequations}\label{eq:determining_new_M}
\begin{widetext}
    \begin{align}
         M^{(1)}_{\alpha, u, u} = \frac{(E^{(1)}_\alpha-\mu_v)}{J_{v, u}^2} \bigg[(E^{(1)}_\alpha-\mu_v) M_{\alpha, v, v}^{(1)} -\sum_{v' \in \mathcal{N}_v \setminus\{u\}} \bigg(J_{v, v'}M_{\alpha, v, v'}^{(1)} + J_{v, v'}^* M_{\alpha, v', v}^{(1)}\bigg)\bigg] + \sum_{v_1', v_2' \in \mathcal{N}_v \setminus \{u\}} \frac{J_{v,v_1'}J_{v, v_2'}^*}{J_{v, u}^2} M_{\alpha, v_1', v_2'}^{(1)} 
    \end{align}
    \end{widetext}
Furthermore, multiplying Eq.~\ref{eq:rewrite_main_eigvals}(a) by $\langle l_\alpha^{(1)}| a_v \ket{G}$ and Eq.~\ref{eq:rewrite_main_eigvals}(b) by $\langle G | a_v |r_\alpha^{(1)}\rangle$ allows us to determine $M_{\alpha, u, v}^{(1)}$ and $M_{\alpha, v, u}^{(1)}$ respectively via
\begin{align}
    &M_{\alpha, u, v}^{(1)}  \nonumber\\
    & = \frac{1}{J_{v, u}}\bigg((E_\alpha^{(1)} - \mu_v) M_{\alpha, v, v}^{(1)} - \sum_{v' \in \mathcal{N}_v \setminus\{u\}} J_{v, v'} M_{\alpha, v', v}^{(1)}\bigg), \\
    &M_{\alpha, v, u}^{(1)}  \nonumber \\
    & = \frac{1}{J_{v, u}}\bigg((E_\alpha^{(1)} - \mu_v) M_{\alpha, v, v}^{(1)} - \sum_{v' \in \mathcal{N}_v\setminus \{u\}} J_{v, v'}^* M_{\alpha, v, v'}^{(1)}\bigg).
\end{align}
\end{subequations}
It can be noted that the right hand sides of all the equations in Eqs.~\ref{eq:determining_new_M}(a), (b) and (c) can be evaluated since $J_{v, u}$ is known, and $v$ as well the vertices in $\mathcal{N}_v \setminus \{u\}$ are in $\mathcal{I}$.

Next, we show that the vertex $u$ can now be added to the set of infected vertices $\mathcal{I}$ --- for this, we need to compute $M^{(1)}_{\alpha, u, w}$ and $M_{\alpha, w, u}^{(1)}$ for all $w \in \mathcal{I}$. This can be done by utilizing the following simple cyclic relationship --- consider a vertex $w \in \mathcal{I}$, the vertex $v \in \mathcal{I}$ which is a neighbour of $u$ and the vertex $u$, then
\[
M^{(1)}_{\alpha, u, w} = \frac{M_{\alpha, u, v}^{(1)} M_{\alpha, v, w}^{(1)}}{\big(M_{\alpha, v, v}^{(1)}\big)^*} \text{ and }M^{(1)}_{\alpha, w, u} = \frac{M_{\alpha, w, v}^{(1)} M_{\alpha, v, u}^{(1)}}{M_{\alpha, v, v}^{(1)}}.
\]
Since $M_{\alpha, v, w}^{(1)}, M_{\alpha, w, v}^{(1)}$ and $M_{\alpha, v, v}^{(1)}$ are known, this allows us to determine $M^{(1)}_{\alpha, u, w}$ for all $w \in \mathcal{I}$. Now, the vertex $u$ can also treated as an infected vertex and added to the set $\mathcal{I}$. Then, using Eq.~\ref{eq:rewrite_main_eigvals}, we can compute the single-particle couplings between $u$ and other sites in $\mathcal{I}$.

Next, we repeat this entire process with another uninfected vertex that can be infected by $\mathcal{I}$ --- our assumption on the graph is that they can be entirely infected starting from the initial $\mathcal{I} = \mathcal{V}_0$ and therefore all the Hamiltonian parameters can be reconstructed.

\subsection{Reconstructing $\chi$}
The second stage involves working in two-photon space, and using the determined single-particle Hamiltonian $H^{(1)}$ along with the measured values of $M^{(2)}_{\vec{\alpha}, v, u}$, where $\vec{\alpha} = (\alpha_1, \alpha_2, \alpha_3)$ and $v, u \in \mathcal{V}_0$, to calculate $\chi_v \ \forall \ v \in \mathcal{V}$ and hence mapping the full Hamiltonian $H$. As with the reconstruction of the single-particle parameters, we will develop a recursive algorithm to reconstruct $\chi_v$ for graphs in which all the vertices can be infected by the vertices at the boundary --- the recursion is more naturally formulated in terms of the coefficients $C^{(2)}_{\vec{\alpha}, v, u}, Q_{\vec{\alpha}, v, u}^{(2)}, R_{\vec{\alpha},v, u}^{(2)}$, which are defined by
\begin{figure*}[t]
	\centering
	\includegraphics[width=0.8\linewidth]{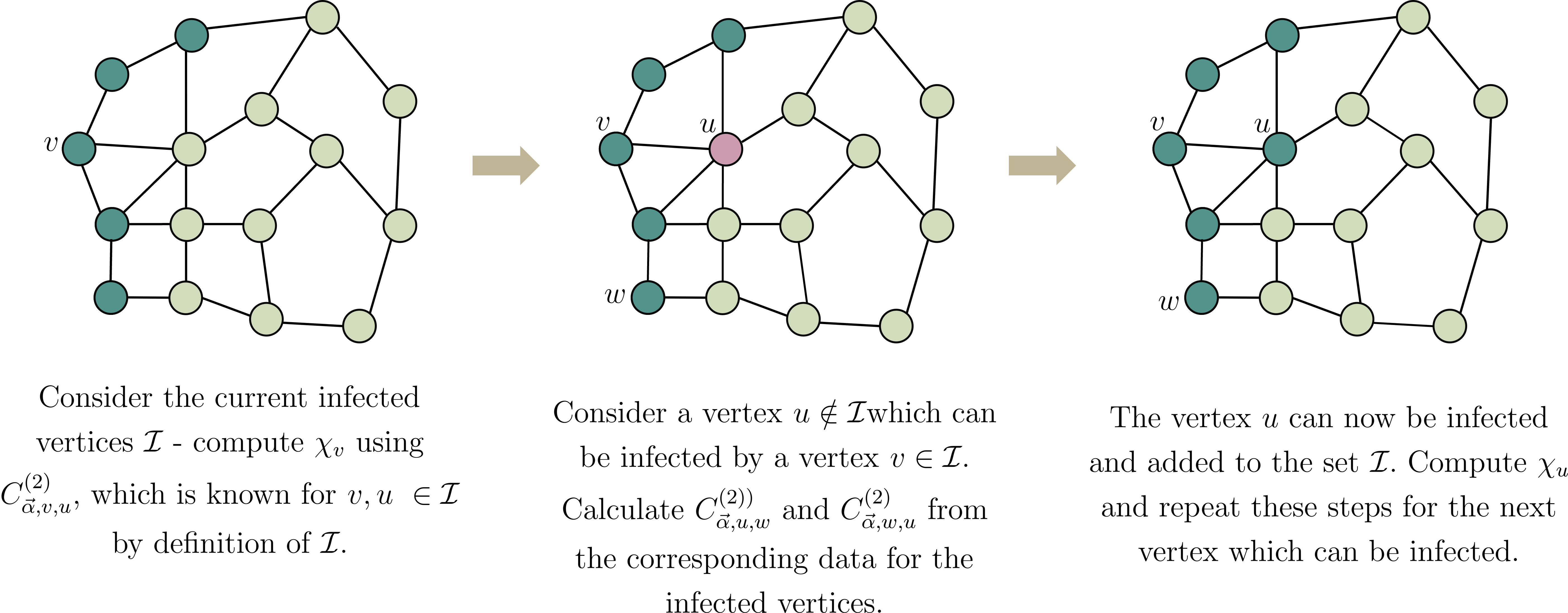}
	\caption{\label{fig:steps_chi} Schematic depicting the tomography algorithm in the two excitation space. The dark-green vertices are the currently infected vertices, while the red vertex is the vertex being infected.}
\end{figure*}
\begin{align*}
&C^{(2)}_{\vec{\alpha}, v, u} = \bra{l_{\alpha_1}^{(1)}} a_v\ket{r_{\alpha_2}^{(2)}} \bra{l_{\alpha_2}^{(2)}}a_u^\dagger \ket{r_{\alpha_3}^{(1)}}, \\
&Q^{(2)}_{\vec{\alpha}, v, u} = \bra{l_{\alpha_1}^{(1)}} a_v^\dagger a_v^2 \ket{r_{\alpha_2}^{(2)}} \bra{l_{\alpha_2}^{(2)}}a_u^\dagger \ket{r_{\alpha_3}^{(1)}}, \\
&R^{(2)}_{\vec{\alpha}, v, u} = \bra{l_{\alpha_1}^{(1)}} a_v \ket{r_{\alpha_2}^{(2)}} \bra{l_{\alpha_2}^{(2)}} (a_u^\dagger)^2 a_u \ket{r_{\alpha_3}^{(1)}}.
\end{align*}
We note that if the single-particle parameters, $J_{v, u}$ and $\mu_v$, have already been reconstructed which in turn allows us to compute the single particle eigenstates $\ket{r_\alpha^{(1)}}, \ket{l_{\alpha}^{(1)}}$, then $C_{\vec{\alpha}, v, u}^{(2)}, Q_{\vec{\alpha}, v, u}^{(2)}, R_{\vec{\alpha}, v, u}^{(2)}$, for $v, u \in \mathcal{V}_0$, can be determined from $M_{\vec{\alpha}, v, u}^{(2)}$ obtained from the boundary measurements. It is straightforward to see this for $C^{(2)}_{\vec{\alpha}, v, u}$ since
\[
C^{(2)}_{\vec{\alpha}, v, u} = \frac{M^{(2)}_{\vec{\alpha}, v, u}}{\bra{G} a_v \ket{r_{\alpha_1}^{(1)}} \bra{l_{\alpha_3}^{(1)}} a_u^\dagger \ket{G}}.
\]
Furthermore, we use the completeness of $\ket{r_\alpha^{(1)}}, \ket{l_{\alpha}^{(1)}}$ in the single-excitation subspace to express $Q_{\vec{\alpha}, v, u}^{(2)}$ and $R_{\vec{\alpha}, v, u}^{(2)}$ in terms of $C_{\vec{\alpha}, v, u}^{(2)}$ via
\begin{subequations}\label{eq:qR_C}
\begin{align}
&Q_{\vec{\alpha}, v, u}^{(2)} = \sum_{\alpha_1'} \bra{l_{\alpha_1}^{(1)}} a_v^\dagger a_v \ket{r_{\alpha_1'}^{(1)}} C_{(\alpha_1', \alpha_2, \alpha_3), v, u}^{(2)}, \\
&R_{\vec{\alpha}, v, u}^{(2)} = \sum_{\alpha_2'} \bra{l_{\alpha_3'}^{(1)}}a_u^\dagger a_u \ket{r_{\alpha_3}^{(1)}} C^{(2)}_{(\alpha_1, \alpha_2', \alpha_3), v, u}.
\end{align}
\end{subequations}

 Similar to the previous subsection and as shown in Fig.~\ref{fig:steps_chi}, at any stage of the reconstruction algorithm, we define the set of \emph{infected vertices} $\mathcal{I}$ as the vertices $v, u$ such that $C_{\vec{\alpha}, v, u}^{(2)}$ and $C_{\vec{\alpha}, u, v}^{(2)}$ are known for all $v, u \in \mathcal{I}$ and for all $\vec{\alpha}$. At the start of the algorithm, $\mathcal{I}$ coincides with the boundary vertices $\mathcal{V}_0$. To develop the reconstruction algorithm, we will use the eigenvalue equation for $H_\text{eff}$ in the two-excitation subspace: $H_\text{eff}\ket{r_{\alpha_2}^{(2)}} = E_{\alpha_2}^{(2)} \ket{r_{\alpha_2}^{(2)}}$ and $\bra{l_{\alpha_2}^{(2)}} H_\text{eff} = E_{\alpha_2}^{(2)} \bra{l_{\alpha_2}^{(2)}}$. Multiplying the right eigenvector equation by $\bra{l_{\alpha_1}^{(1)}} a_v$ and the left eigenvector equation by $a_v^\dagger \ket{r_{\alpha_3}^{(1)}}$, we obtain
\begin{subequations}\label{eq:two_particle_eigval}
\begin{align}
    & \chi_v \bra{l^{(1)}_{\alpha_1}} a_v^\dag a_v^2\ket{ r^{(2)}_{\alpha_2}}  +\sum_{v' \in \mathcal{N}_v} J_{v, v'} \bra{ l^{(1)}_{\alpha_1}} a_{v'} \ket{ r^{(2)}_{\alpha_2} } =   \notag \\
   & \quad\qquad\qquad (E^{(2)}_{\alpha_2} - \mu_v - E^{(1)}_{\alpha_1} ) \langle l^{(1)}_{\alpha_1} | a_v | r^{(2)}_{\alpha_2} \rangle  = 0, \\
    &\chi_v \langle l^{(2)}_{\alpha_2} | a_v^{\dag 2} a_v | r^{(1)}_{\alpha_3} \rangle +  \sum_{v' \in \mathcal{N}_v} J_{v', v}\langle l^{(2)}_{\alpha_2} | a^\dag_{v'} | r^{(1)}_{\alpha_3} \rangle +  \notag \\
   & \quad\qquad\qquad  (E^{(2)}_{\alpha_2} - \mu_v - E^{(1)}_{\alpha_3}) \langle l^{(2)}_{\alpha_2} | a^\dag_v | r^{(1)}_{\alpha_3} \rangle  = 0.
\end{align}
\end{subequations}
From these eigenvalue equations, we can now obtain $\chi_v$ for $v \in \mathcal{I}$. First, we note that since $\ket{r_\alpha^{(1)}}, \ket{l_\alpha^{(1)}}$ and $\ket{r_\alpha^{(2)}}, \ket{l_\alpha^{(2)}}$ are complete within the first and second excitation subspaces respectively, 
\[
\sum_{\alpha_1, \alpha_2}  \langle l^{(1)}_{\alpha_1} | a_v | r^{(2)}_{\alpha_2} \rangle \langle l^{(2)}_{\alpha_2} | a^{\dag}_u | r^{(1)}_{\alpha_1} \rangle = 2\delta_{v, u}.
\]
Then, multiplying Eq.~\ref{eq:two_particle_eigval}a by $\bra{l_{\alpha_2}^{(2)}} a_u^\dagger \ket{r_{\alpha_1}^{(1)}}$ and summing over the indices $\alpha_1, \alpha_2$, we obtain that
\begin{align}\label{eq:findingchi_v}
      \chi_v  = -\frac{1}{2}\sum_{\alpha_1,\alpha_2} (\mu_v + E^{(1)}_{\alpha_1} - E^{(2)}_{\alpha_2}) C^{(2)}_{(\alpha_1, \alpha_2, \alpha_1),v,v}.
 \end{align}
If $v \in \mathcal{I}$, then the right hand side of Eq.~\ref{eq:findingchi_v} is completely known, and thus $\chi_v$ can be reconstructed.

Next, we consider a vertex $u \in \mathcal{V}\setminus \mathcal{I}$ and can be infected by a vertex in $\mathcal{I}$. By definition, this implies that $u$ is the \emph{unique uninfected nearest neighbour} of a vertex $v \in \mathcal{I}$ [Fig.]. In what follows, we show that the vertex $u$ can be added to the set of infected vertices $\mathcal{I}$ since we can determine $C_{\vec{\alpha}, w, u}^{(2)}$ and $C_{\vec{\alpha}, u, w}^{(2)}$ for all $w \in \mathcal{I}$. First, consider determining $C_{\vec{\alpha}, u, w}^{(2)}$ --- this can be done multiplying Eq. \ref{eq:two_particle_eigval}(a) by $ \bra{ l^{(2)}_{\alpha_2} } a^{\dag}_w | r^{(1)}_{\alpha_3}\rangle$ to obtain:
 \begin{gather*}
     C_{\vec{\alpha}, u, w}^{(2)} = \frac{1}{J_{v,u}} \biggl(\big( E^{(2)}_{\alpha_2} - \mu_v - E^{(1)}_{\alpha_1}\big)C_{\vec{\alpha},v,w}^{(2)} - \nonumber\\
     \qquad \sum_{v' \in \mathcal{N}_v \setminus \{u\}} J_{v,v'} C_{\vec{\alpha},v',w}^{(2)} - \chi_v Q_{\vec{\alpha},v,w}^{(2)} \biggr).
 \end{gather*}
Observe that, from this equation, the evaluation of $C_{\vec{\alpha}, u, w}^{(2)}$ requires the coefficients $C_{\vec{\alpha}, v_1, v_2}^{(2)}, Q_{\vec{\alpha}, v_1, v_2}^{(2)}$ for only infected vertices $v_1, v_2 \in \mathcal{I}$ --- $C_{\vec{\alpha}, v_1, v_2}^{(2)}$ is known by definition, and $Q_{\vec{\alpha}, v_1, v_2}^{(2)}$ can be obtained using Eq.~\ref{eq:qR_C}. A similar procedure can be followed to obtain $C_{\vec{\alpha}, w, u}^{(2)}$ --- multiplying Eq.~\ref{eq:two_particle_eigval}b by $\bra{l_{\alpha_1}^{(1)}} a_w \ket{r_{\alpha_2}^{(2)}}$, we obtain 
 \begin{gather*}
    C_{\vec{\alpha},  w, u}^{(2)} = \frac{1}{J_{u, v}} \biggl(\big( E^{(2)}_{\alpha_2} - \mu_v - E^{(1)}_{\alpha_1}\big)C_{\vec{\alpha},w, v}^{(2)} - \nonumber\\
     \qquad \sum_{v' \in \mathcal{N}_v \setminus \{u\}} J_{v', v} C_{\vec{\alpha},w, v'}^{(2)} - \chi_v R_{\vec{\alpha},w, v}^{(2)} \biggr),
 \end{gather*}
 which gain involves only coefficients $C_{\vec{\alpha}, v_1, v_2}^{(2)}$ and $R_{\vec{\alpha}, v_1, v_2}^{(2)}$ corresponding to infected vertices $v_1, v_2 \in \mathcal{I}$, and thus can be entirely determined using the known $C_{\vec{\alpha}, v_1, v_2}^{(2)}$ and Eq.~\ref{eq:qR_C}. Since we have successfully determined both $C_{\vec{\alpha}, u, w}^{(2)}$ and $C_{\vec{\alpha}, w, u}^{(2)}$ for all $w \in \mathcal{I}$, we can add the vertex $u$ to the set of infected vertices. Then, we can go onto determine $\chi_u$ using Eq.~\ref{eq:findingchi_v} and repeat this process until all the vertices have been infected, and thus all the on-site anharmonicities have been determined.

\subsection{Impact of measurement errors}\label{sec:stability}
In the last two subsections, our analysis assumed that the measurements are perfect i.e.~the boundary data $E_\alpha^{(1)}, E_\alpha^{(2)}, M_{\alpha, v, u}^{(1)}$ and $M_{\vec{\alpha}, v, u}^{(2)}$ are known perfectly. However, in an experimentally realistic setting, there will be an error in the measured data --- this raises the question of how much the reconstructed Hamiltonian parameters diverge from true ones when measurements contain errors. The primary inputs into the algorithm are the eigenenergies $E_\alpha^{(1)}, E_\alpha^{(2)}$ and the coefficients $M^{(1)}_{\alpha, v, u}, M_{\vec{\alpha}, v, u}^{(2)}$ (or alternatively $C_{\vec{\alpha}, v, u}^{(2)}$). In a well-calibrated experimental setup, it is reasonable to assume that the eigen-energies can be determined to an accuracy that is much higher than the eigenvector overlaps since the eigenenergies can be determined entirely through the position of the resonances. Hence, in this section, we focus on determining scaling of the error in the reconstructed couplings with the error in the coefficients. 

\begin{figure*}[t]
	\centering
	\includegraphics[width=0.75\linewidth]{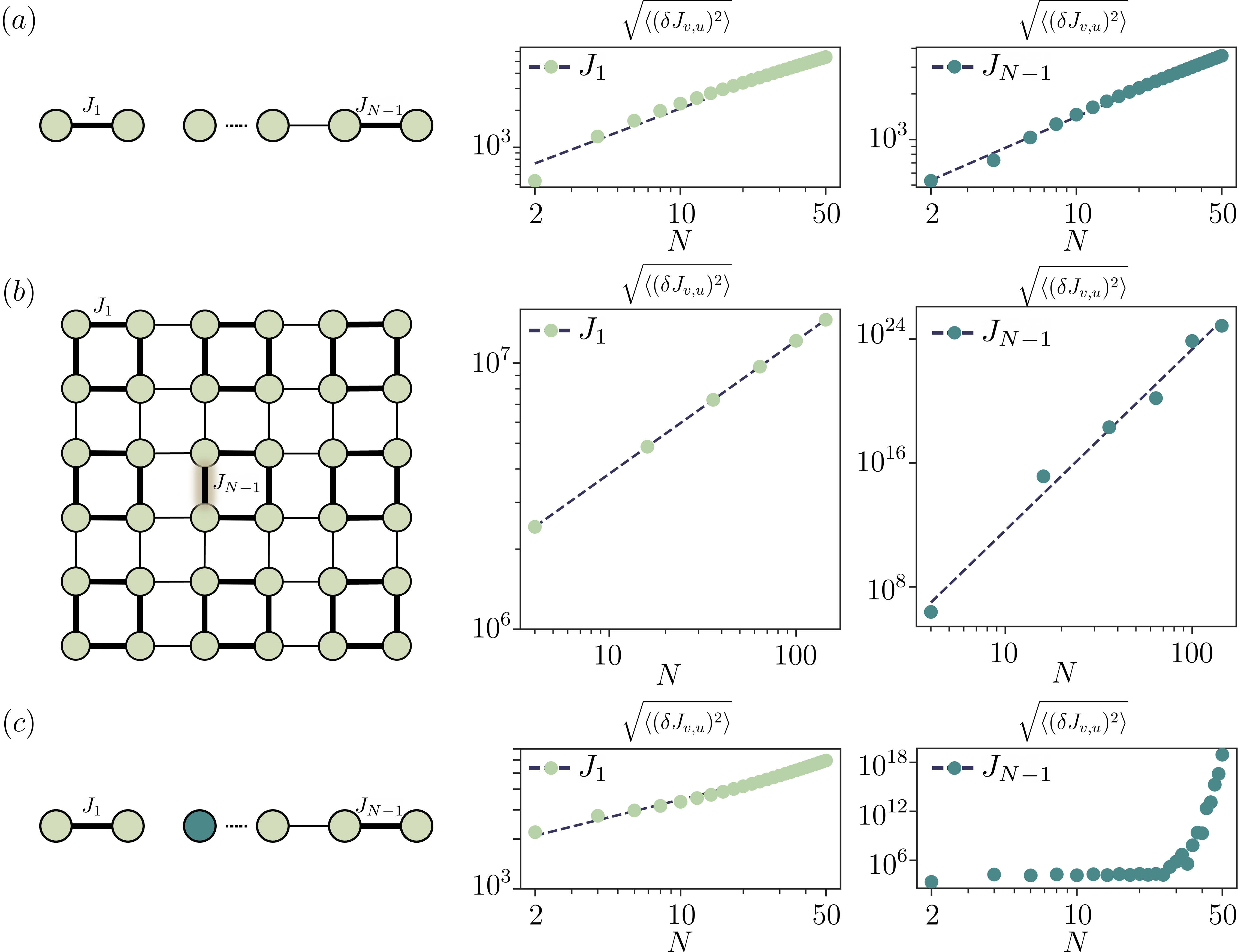}
	\caption{\label{fig:stab} Stability analysis. The $x$-axis denotes number of sites in the array on a log scale, y-axis denotes the error metric $\sqrt{\langle (\delta J_{v, u})^2\rangle} $ on a log scale.(a) Schematic depicting a 1D SSH. Error metric plot for a SSH chain with $J_1/J_2 = 1.2$. (b) Schematic depicting a 2D SSH. Error metric plot for a SSH $2D$ lattice with $J_1/J_2 = 1.2$. Each point is in $100$ point average with disorder in onsite potentials to be $10\ MHz$ to avoid absolute degeneracies. (c) Schematic depicting a 1D SSH with a defect site at $N/2$ s.t. $\mu_{N/2} = \mu_{N/2} + 2J_1$. Error metric plot s.t. $J_1/J_2 = 1.2$. As evident, error metric for $J_{N-1}$ depicts a breakdown in stability.}
\end{figure*}

To illustrate how the error scales with increasing $N$ we take up the Su-Schrieffer-Heeger (SSH) model \cite{su1979solitons} which can described by the Hamiltonian:
\begin{align*}
% \label{hamiltonian_SSH}
	H_\text{SSH} =\sum_{v} \bigg(J_1 a_{2v + 1}^\dag a_{2v} + J_2 a_{2v + 2}^\dag a_{2v + 1} + \text{h.c.}\bigg),
\end{align*}
where each site of the lattice has the same onsite potential $\mu_0$, but the couplings alternate between $J_1, J_2$. For a small error $\delta M_{\alpha, v, u}^{(1)}$, $\delta C_{\vec{\alpha}, v, u}^{(2)}$, the errors in the reconstructed coefficients $J_{v, u}$ and $\chi_v$ can be approximated by
\begin{align*}
% \label{PD}
      &\delta J_{v,u} = \sum_{\alpha, v', u'} \frac{\partial J_{v,u}}{\partial M^{(1)}_{\alpha, v', u'}} \delta M^{(1)}_{\alpha, v', u'}, \\
      &\delta \chi_v = \sum_{\vec{\alpha}, v, u} \bigg(\frac{\partial \chi_{v}}{\partial C^{(2)}_{\vec{\alpha}, v', u'}} \delta C^{(2)}_{\vec{\alpha}, v', u'} + \frac{\partial \chi_v}{\partial J_{v', u'}}\delta J_{v', u'}\bigg).
 \end{align*}
Further we assume that errors in the coefficients are uncorrelated i.e. $\langle \delta M^{(1)}_{\alpha, v, u} \delta M^{(1)}_{\alpha', v', u'} \rangle = \delta_{\alpha, \alpha'} \delta_{v, v'}\delta_{u, u'}$, $\langle \delta C^{(2)}_{\vec{\alpha}, v, u} \delta C^{(2)}_{\vec{\alpha'}, v', u'} \rangle = \delta_{\vec{\alpha}, \vec{\alpha'}} \delta_{v, v'}\delta_{u, u'}$ and $\langle \delta M^{(1)}_{\alpha, v, u} \delta C^{(2)}_{\vec{\alpha}, v, u}\rangle = 0$, which is a reasonable assumption to make since we are dealing with errors at different frequencies, we can then obtain that
\begin{align*}
&\langle (\delta J_{v, u})^2\rangle = \sum_{\alpha, v', u'} \bigg | \frac{\partial J_{v, u}}{\partial M_{\alpha, v', u'}^{(1)}}\bigg |^2, \text{ and }\\
&\langle (\delta \chi_{v})^2\rangle = \sum_{v', u'} \bigg(\sum_{\alpha}\bigg | \frac{\partial \chi_{v}}{\partial M_{\alpha, v', u'}^{(1)}}\bigg |^2 + \sum_{\vec{\alpha}}\bigg | \frac{\partial \chi_{v}}{\partial C_{\vec{\alpha}, v', u'}^{(2)}}\bigg |^2\bigg)
\end{align*}
In Fig. \ref{fig:stab}(a) we take a SSH chain with real coupling strengths and plot the behavior of the defined error metric for the coupling strengths along the first and last edge in the chain. From the plot, we can see that the metric scales polynomially with $N$. This suggests that even though the amount of error in the reconstructed coupling strengths increase with $N$, we only need a polynomial increase in the number of measurements to accurately determine the Hamiltonian in question. We show the same for a $2D$ SSH-like model in   Fig. \ref{fig:stab}(b). We expect this linear scaling of error (on a log scale) for models with delocalized eigenvectors. If, however, we have a localized state, then to accurately determine the Hamiltonian parameters in the interior of the lattice may require exponentially accurate measurements. For example, consider a SSH chain with at least one strongly localized eigenvector obtained by inducing a defect site in the middle such that the onsite potential of this site is greater than two times the maximum coupling in the chain. This localized eigenvector will be rapidly decay away from the defect site as we approach the edge site at which measurements can be performed. Hence, we expect that in this case there will exist a $N$ after which a small error in the measurement of coefficients will lead to a blow up in the error of the estimated coupling strengths. We plot this case in Fig. \ref{fig:stab}(c). We confirm from the plot that after a specific $N$ the error metric scales polynomially implying that an exponentially large number of measurements will be needed to accurately determine the realized Hamiltonian after this point.

%  \begin{figure}[!htbp]
% 	\centering
% 	\includegraphics[width=0.45\linewidth]{Fig9.png}
% 	 % \setlength{\abovecaptionskip}{-10pt}
% 	 % \setlength{\belowcaptionskip}{-10pt}
% 	\caption{\label{fig:fig1009} Stability analysis. x-axis denotes number of sites in the array on a log scale, y-axis denotes the error metric on a log scale. Top: Schematic depicting the measurement scheme for 1D SSH with a defect site at $N/2$ s.t. $\mu_{N/2} = \mu_{N/2} + 2J_1$. Error metric plot s.t. $J_1/J_2 = 1.2$. As evident, error metric for $J_{N-1}$ depicts a breakdown in stability.}
% \end{figure}

% (see Appendix \ref{apxA}).\\
 
\begin{figure}[b]
	\centering
	\includegraphics[width=0.65\linewidth]{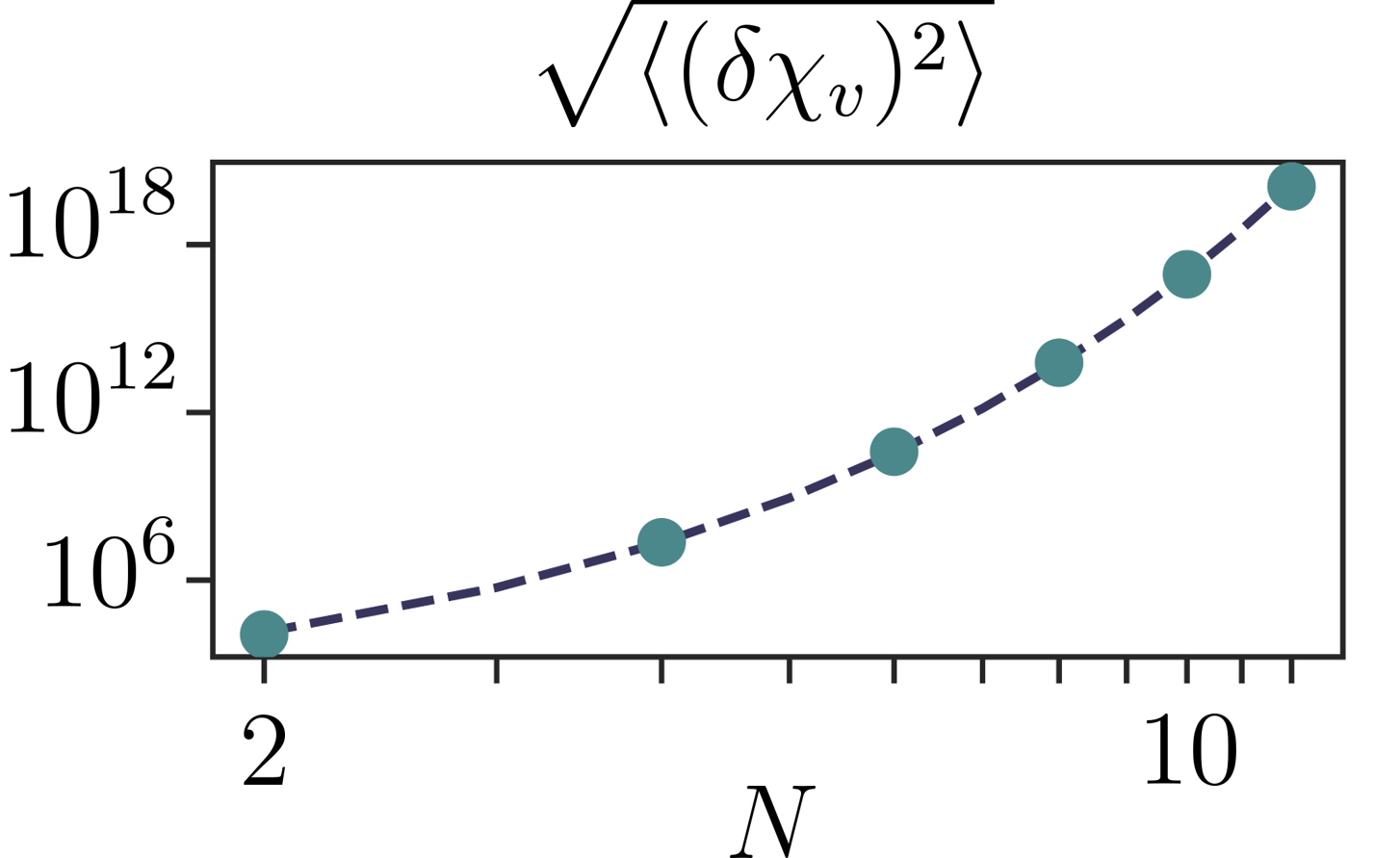}
	\setlength{\belowcaptionskip}{-10pt}
	\caption{\label{fig:chi}  Stability analysis where x-axis (log scale) denotes number of sites in the array and y-axis (log scale) denotes the error metric $\sqrt{\langle (\delta \chi_{v})^2\rangle}$ in $\chi_N$ for a SSH chain with $J_1/J_2 = 1.2$. $\sqrt{\langle (\delta \chi_{v})^2\rangle}$ scales polynomially with $N$ on a log scale. We assume $\delta J = 0$}
\end{figure}

 Lastly, we move onto the second-stage of the tomography scheme, involving reconstruction of $\chi$s. We again use a similar 1D SSH chain and plot the behavior of $\sqrt{\langle (\delta \chi_{v})^2\rangle}$ in Fig. \ref{fig:chi} assuming zero error in reconstruction of couplings as the number of sites $N$ increases. We notice that the error metric scales polynomially with $N$ on a log scale, which implies that reconstructing non-linear part of the Hamiltonian requires exponentially large number of measurements as the number of sites $N$ increases. This indicates that this algorithm despite being applicable for lossy lattices with complex hopping rates is only practical when the size of the photonic lattice (characterized by $N$) is small. 
 Now, however, if we assume an additional capability in the system under investigation, such that the onsite non-linearity can be toggled on/off during the experiment at all nodes, there exists a modification of the algorithm that remains stable irrespective of the system size $N$. The assumption of control over onsite non-linearity is generally valid for most on-chip typical quantum simulators, as local tunability is a necessity in order to precisely program the terms of the Hamiltonian being implemented \cite{carusotto2020photonic}. In these systems, obtaining concurrent local measurability with local control is experimentally challenging due to the difficulty in placing impedance matched read-out resonators (for superconducting systems) or lossless waveguides (for photonics) in proximity of the densely packed resonator arrays used to realize these bosonic lattices \cite{kim2021quantum, zhang2023superconducting, owens2022chiral, kollar2019hyperbolic, saxena2022realizing}. Hence, our modification that overcomes this issue can be useful for tomography of non-linearities in such bosonic lattices. Under this assumption, the non-linearity for any site $v \in \mathcal{V}$ can be calculated as:
\begin{align*}
% \label{ch:me:eq10}
        \chi_v = \text{Tr}(H) - \text{Tr}(H_{v,\text{off}}) = \sum_\alpha \left( E^{(2)}_{\alpha} - E^{(2)}_{\alpha, v,\text{off}} \right)
\end{align*}
where $E^{(2)}_{\alpha, \text{off}}$ denotes the eigenenergies in two-photon subspace of the lattice measured at the boundaries with non-linearity at site $v$: $\chi_v$ turned off, $H_{v,\text{off}}$ denotes the corresponding system Hamiltonian.

\section{Quantum enhancement protocol}\label{sec:quantum_enhancement}
The algorithm described in section \ref{sec:tomography_with_tran_g2}, while in several experimentally relevant settings allowing us to stably reconstruct the Hamiltonian parameters, operates at the standard quantum limit (SQL). More precisely, suppose our goal is to reconstruct the Hamiltonian parameters to a precision $\varepsilon$ --- then, the precision required in the boundary measurements (either transmission or two-photon correlation) is also $\varepsilon_\text{bd} = \Theta(\varepsilon)$. Consequently, even in an ideal setting of no experimental measurement errors, the number of photons (used as independent measurement trials) needed to achieve this target precision scales as $1/\varepsilon_\text{bd}^2 \sim O(1/\varepsilon^2)$, which is the well known standard quantum limit. We point out that $\varepsilon_\text{bd}$ can either scale polynomially with $N$ or exponentially/superpolynomially with $N$ depending on the spectral properties of the Hamiltonian in question (for e.g.~see the discussion in section \ref{sec:stability}) --- throughout this section, we will focus on the dependence of $\varepsilon_\text{bd}$ on the target precision $\varepsilon$ and suppress the dependence of $\varepsilon_\text{bd}$ on $N$.

A natural question to ask is if the dependence on the number of measurements on $\varepsilon$ can be improved to beyond $1/\varepsilon^2$ using techniques from quantum-enhanced sensing. In this section, we address this question, and adapt techniques from quantum-enhanced interferometry to obtain a tomography protocol which assumes (a) boundary excitation and measurement access and (b) ability to toggle the nonlinear term in the Hamiltonian, and beats the standard quantum limit. We point out that, while compared to the tomography method described in section \ref{sec:tomography_with_tran_g2}, we additionally require the ability to switch on and off the nonlinear term in the Hamiltonian for the protocol described in this section, this requirement is satisfied in many experimental systems \cite{kim2021quantum, zhang2023superconducting,owens2022chiral}, and introducing this ability is typically experimentally much easier than being able to excite and measure at a site in the bulk of the lattice  \cite{kim2021quantum, zhang2023superconducting, owens2022chiral, kollar2019hyperbolic, saxena2022realizing}.
%Finally, it can then be asked if we can achieve an improvement in this precision by utilizing non-classical states of photons. In the next section we show that, if we impose a few additional restrictions on the system being studied, we can reduce the rate in scaling of photons needed in comparison to the classical interferometry to obtain an $\Delta$ precision in measurement by utilizing quantum mechanical enhancement due to the entangled nature of the input photons. Particularly, we make the assumptions that the lattices in question are lossless, we have control over the number of input/output ports of the system and, that the onsite non-linearity can be arbitrarily turned on/off during the experiment at all nodes. Then, our proposition for this entanglement enhanced protocol is that the Hamiltonian parameters $J_{v,u}$, $\mu_v$ and $\chi_v$ for a fixed number of sites can be obtained with a precision $\Delta$ from just boundary quantum interferometry measurements using $N00N$ states of $O(\Delta^{-2/3})$ photons.
\subsection{Reconstructing $\mu_v$ and $J_{v, u}$}
\begin{figure}
    \centering
    \includegraphics[width=1\linewidth]{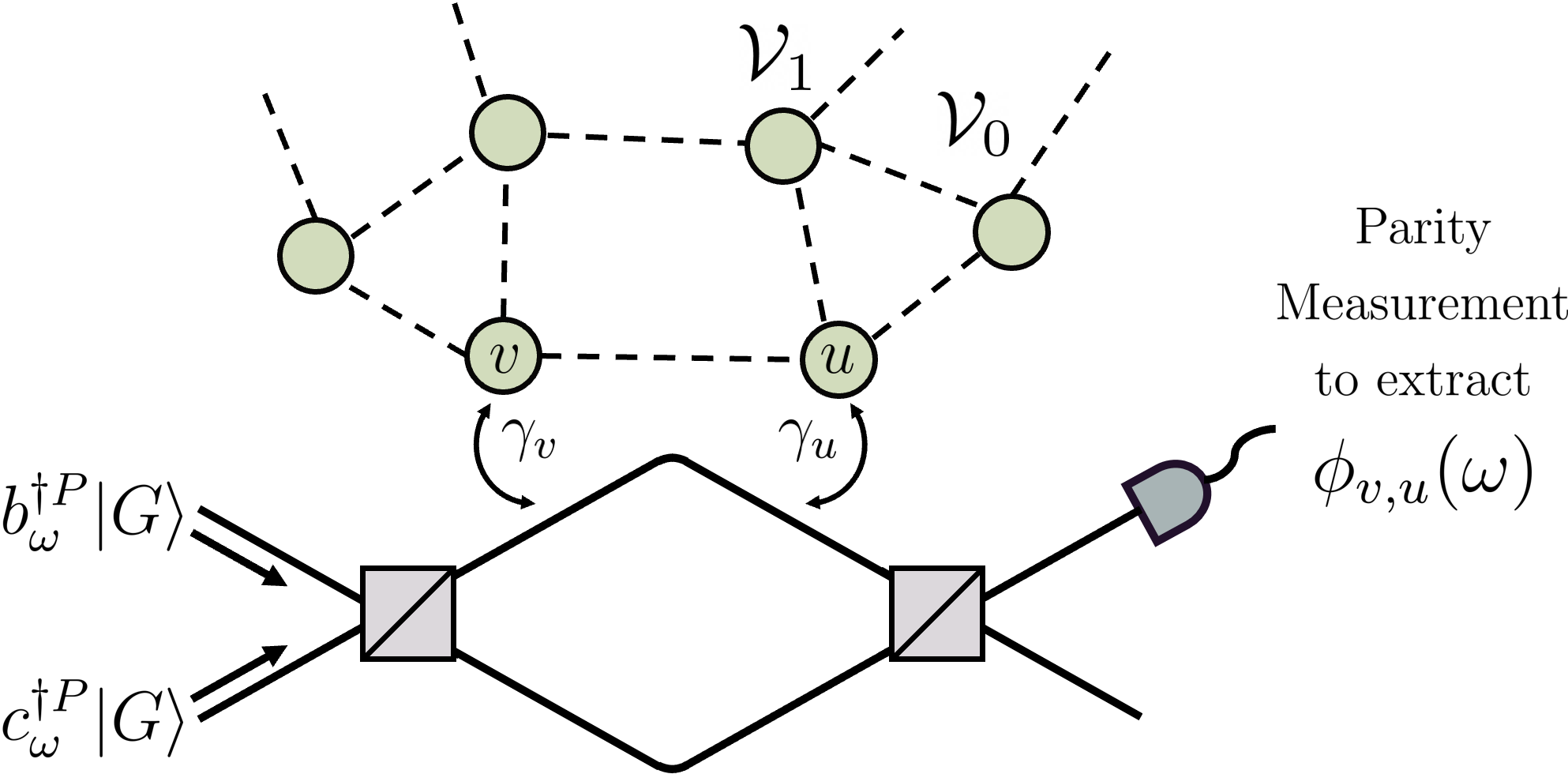}
    \caption{Obtaining a quantum enhancement in the measurement of the single-particle parameters $J_{v, u}$ and $\mu_v$ by using it as the phase-shifting element in the standard fock state spectroscopy setup.}
    \label{fig:enhancement_1}
\end{figure}

Consider first determining the single-particle parameters $J_{v, u}$ and $\mu_v$ --- here, the number of photons needed to sense these parameters can be quadratically improved by using the standard Fock-state interferometry \cite{ bollinger1996optimal, holland1993interferometric}. More precisely, instead of using $N$ single photons to perform the transmission measurement, we can instead use the setup depicted in Fig.~\ref{fig:enhancement_1} --- boundary sites $v, u \in \mathcal{V}_0$ are coupled to one arm of a Mach-Zender interferometer (MZI) and throughout this measurement, we turn-off the non-linear term in the Hamiltonian. The lattice effectively acts as a (frequency-dependent) phase shifter, with the phase imparted to the MZI arm given by
\[
\phi_{v, u}(\omega) = \text{arg}\bigg(1 + i\sum_{\alpha, u, v} \frac{\sqrt{\gamma_u \gamma_v} M_{\alpha,u,v}^{(1)}}{\omega-E_\alpha}\bigg)
\]
Clearly, measuring $\phi_{v, u}(\omega)$ allows us to determine the which is then excited with continuous-wave $P$-photon Fock states in both of its arms.
A parity measurement on the output port of the MZI allows us to measure the phase imparted to the photons due to scattering from the boundary site $v$ to a precision of $O(P^{-1})$ \cite{ bollinger1996optimal, holland1993interferometric}, which in-turn allows us to determine $M_{\alpha, v, u}^{(1)}$ to a precision $\varepsilon_\text{bd} = O(P^{-1})$. Applying the algorithm described in section \ref{sec:tomography_with_tran_g2} therefore allows us to determine $\mu_v$ and $J_{v, u}$ to precision $\varepsilon$ with $O(\varepsilon^{-1})$ photons, thus yielding a quadratic improvement over the standard quantum limit.

\subsection{Reconstructing $\chi_v$}
Obtaining a quantum enhancement in precision of non-linearities $\chi_v$'s is more challenging than simply using the Fock state spectroscopy setup since in the presence of the on-site non-linearity which in turn induces photon-photon interactions, an incident Fock state is scattered into a complex multi-mode state from which extracting $\chi_v$ is not straightforward. Instead, by carefully turning on and off the non-linearity in the lattice, we describe below a scheme that can still be used to obtain a quantum enhancement.

\begin{figure}
    \centering
    \includegraphics[width = 1\linewidth]{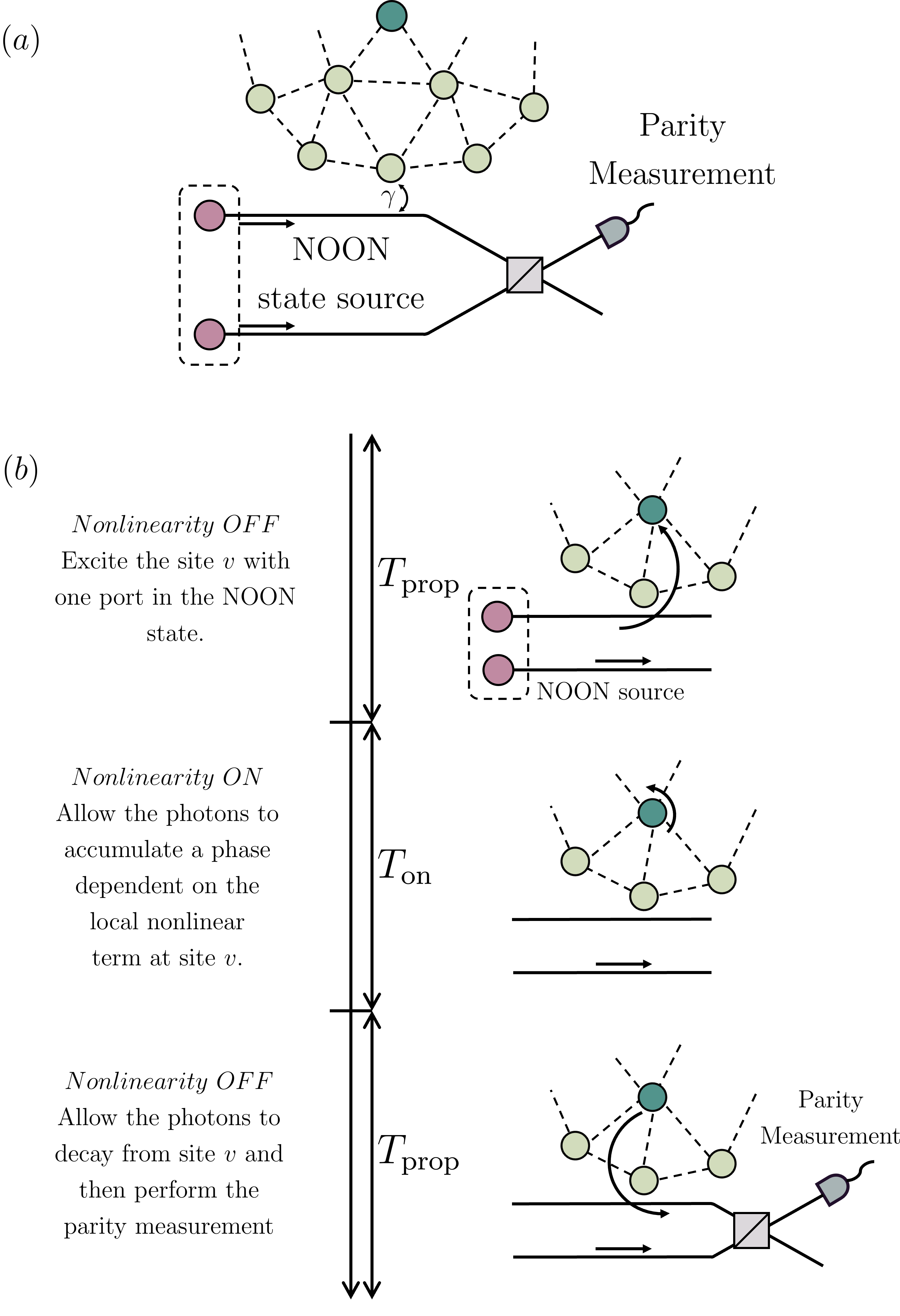}
    \caption{(a) Setup depicting the use of $NOON$ states in obtaining quantum enhancement in extracting the required Hamilotnian parameters. (b) The proposed parity measurement scheme. $T_{\text{prop}}$ denotes the propagation time for photons to reach from the coupling waveguide to the site of interest. The non-linearity remain $OFF$ during this interval. $T_{\text{on}}$ denotes the interval where the non-linearity at the sit of interest is turned $ON$.This interval is again followed up with a time interval $T_{\text{prop}}$ which allows to photons to couple back into the input/output waveguide. The required quantities can then obtained by performing a parity measurement.}
    \label{fig:figure_quantum_chi}
\end{figure}

\emph{Protocol}. Suppose we want to estimate non-linearity $\chi_v$ at a site $v \in \mathcal{V}$ --- we adapt the standard NOON state interferometry \cite{sanders1989quantum} and couple a boundary site to one of the two ports that are carrying the NOON state [Fig.~\ref{fig:figure_quantum_chi}(a)].
\emph{First}, with the non-linearity turned off, we excite the boundary sites so as to initialize the system in a superposition of vacuum and $P$ photon Fock state at site $v$ i.e. in the state $\ket{\Psi_{P, v}}$
\begin{align}
\label{eq:NOON_state_ideal}
\ket{\Psi_{P, v}} = \frac{1}{\sqrt{2}}\bigg(\ket{0}\otimes \ket{P}_\text{ref} + e^{i\alpha}\frac{a^{\dag P}_v}{ \sqrt{P!}}\ket{G}\otimes \ket{G}_\text{ref} \bigg),
\end{align}
where $\ket{P}_\text{ref}$ ($\ket{G}_\text{ref}$) refers to the $P$-photon ($0$-photon) state in the reference port not coupled to the lattice, and $\alpha$ is a reference phase that we will pick later.
Since the single-particle parameters have already been measured and the non-linearities are turned off, this can be done entirely with access to the boundary sites. More concretely, we couple the interferometer port to a boundary site [Fig.~\ref{fig:figure_quantum_chi}(a)], and then compute the wave-packet that is emitted into this port if a single particle is initialized at site $v$ and allowed to evolve with respect to the Hamiltonian $H$. Then, the $P$ photons in the mode coupling to the lattice are prepared in the time reversed wave-packet, and the $P$ photons in the other mode in the interferometer is prepared in the emitted wave-packet. This state, when made incident on the lattice, excites $P$ particles at site $v$ and furthermore, when these $P$ photons are emitted back into the port, they can be interfered without a mode-mismatch with the photons in the other port. Appendix \ref{apxC} provides explicit details about the computation of this wave-packet, as well as on how a NOON, in a given wave-packet, can be prepared experimentally.

However, we need to account for the fact that the single-particle parameters are not known exactly from the first stage of the tomography algorithm, and consequently this process, instead of preparing the state $\ket{\Psi_{P, v}}$ in Eq.~\ref{eq:NOON_state_ideal}, prepares a state $\ket{\hat{\Psi}_{P, v}}$, where
\begin{align}
\label{eq:NOON_state_appx}
\ket{\hat{\Psi}_{P, v}} = \frac{1}{\sqrt{2}}\bigg(\ket{G} \otimes \ket{P}_\text{ref} + e^{i\alpha}\ket{\hat{\psi}_{P, v}}\otimes \ket{G}_\text{ref}\bigg),
\end{align}
where $\ket{\psi_{P, v}}$ approximates the state $a_v^{\dag P}\ket{G}/\sqrt{P!}$. In particular, we assume that the coefficients $J_{v, u}$ and $\mu_v$ are determined to a precision $\varepsilon_0$, and that the photons decay into the output port, from any site in the interior, in time approximately $T_\text{prop}$. Equivalently for the time-reversed wave-packet to excite a photon at the target site, it can be shown that for all $v \in \mathcal{V}$ (see Appendix \ref{apxC} for a detailed proof):
\begin{align}\label{eq:error_first_stage_calib}
\norm*{\frac{a_v^{\dag P}}{\sqrt{P!}}\ket{G}-\ket{\hat{\psi}_{P,v}}} \leq \sqrt{T_{\text{prop}} P N (d + 1) \varepsilon_0}.
\end{align}
Note that increasing the number of photons $P$ used for reconstructing $\chi_v$ also increases the error between $a_v^{\dag P} \ket{G} / \sqrt{P!}$ and $\ket{\hat{\psi}_{P,v}}$. Consequently the precision $\varepsilon_0$ of the single-particle parameters, as reconstructed from the first stage of this algorithm, would need to be decreased with $P$ which would imply that the number of photons used in that stage would need to increase with $P$. However, as we argue below, to obtain even a super Heisenberg scaling in precision of $\chi_v$, the overhead needed in terms of the number of photons from the first stage of the algorithm is much less than $P$, and thus does not impact the super Heisenberg scaling obtained for $\chi_v$. 

Next, we turn on the non-linearity $\chi_v$ at site $v$ for a time period $T_\text{on}$, which remains to be chosen. If $T_\text{on}$ is sufficiently short, then the $P$ particles initialized at site $v$ does not diffuse significantly to the neighbouring sites, and we prepare a state
\begin{align}\label{eq:phase_acc_NOON_ideal}
\ket{\hat{\Phi}_{P, v}} = e^{-iHT_\text{on}} \ket{\hat{\Psi}_{P, v}},
\end{align}
which approximates the state $\ket{\Phi_{P, v}}$, where
\begin{subequations}\label{eq:chi_v_bs_equation}
\begin{align}
 \ket{\Phi_{P, v}}=  \frac{1}{\sqrt{2}} \left(\ket{G}\otimes \ket{P}_\text{ref} + e^{i(\alpha - \theta_{P, v})}\frac{a_v^{\dag P}}{\sqrt{P!}}\otimes \ket{G}_\text{ref}\right),
\end{align}
where 
\begin{align}
\theta_{P, v} = \bigg(\mu_v P + \frac{\chi_v P(P - 1)}{2}\bigg) T_\text{on}
\end{align}
\end{subequations}
A more careful analysis of this stage, detailed in appendix \ref{apxC}, allows us to provide the following quantitative upper bound on the error between $\ket{\psi}$ and $\ket{\phi_{P_v}}$:
\begin{align}\label{eq:error_diffusion}
\norm{\ket{\hat{\Phi}_{P, v}} - \ket{\Phi_{P, v}}} \leq \sqrt{P} JdT_\text{on} + \sqrt{T_{\text{prop}} P N (d + 1) \varepsilon_0},
\end{align}
where $d$ is the degree of the graph $(\mathcal{V}, \mathcal{E})$ describing the lattice (i.e. the maximum number of edges incident on any one vertex), and $J$ is an upper bound on the coupling coefficients $J_{v, u}$ (i.e.~$\abs{J_{v, u}} \leq J$ for all $(v, u) \in \mathcal{E}$). This upper bound implies that for short enough $T_\text{on} \leq O(P^{-1/2})$, then the error between $\norm{\ket{\hat{\Phi}_{P, v}} - \ket{\Phi_{P,v}}}$ can be controlled to remain small even when increasing $P$.

Finally, we can now measure the phase $\theta_{P, v}$ introduced in Eq.~\ref{eq:chi_v_bs_equation} by performing the parity measurement after applying a beam splitter operation. Choosing $\alpha = 0$ and $\pi/2$ in the original NOON state yields an expected outcome of the quadratures $\cos(\theta_{P, v})$ and $\sin(\theta_{P, v})$ respectively. These expectation values then allow us to extract $\theta_{P, v}$ --- however, just the probabilistic nature of the measurement yields that there is a constant (i.e.~$P$ independent error) in the measured quadratures. However, since
${d\chi_v}/{d\theta_{P, v}} = {2}/{P( P - 1) T_\text{on}}$, 
being able to measure $\theta_{P, v}$ to a constant precision allows us to measure $\chi_v$ to a precision that is enhanced by a factor of $1 / P(P-1) T_\text{on}$. 

\emph{Sensitivity Analysis}. In the remainder of this subsection, we sketch the sensitivity analysis of the protocol described above --- a rigorous and detailed version of this analysis is provided in appendix \ref{apxC}. In the idealized setting of no experimental errors, there are three sources of errors in our protocol --- \emph{first} is the error in the measured single-particle parameters $J_{v, u}$ and $\mu_v$ determined in the first stage of the protocol and used for designing the incident NOON state's wave-packet. This manifests as an error between $\ket{\hat{\Psi}_{P, v}}$ and $\ket{\Psi_{P, v}}$ defined in Eqs.~\ref{eq:NOON_state_ideal} and \ref{eq:NOON_state_appx} respectively. \emph{Second} is the error arising due to the diffusion of photons into the sites other than the target site $v$, which manifests as an error between $\ket{\hat{\Phi}_{P, v}}$ and $\ket{\Phi_{P, v}}$ defined in Eqs.~\ref{eq:chi_v_bs_equation} and \ref{eq:phase_acc_NOON_ideal}. \emph{Finally}, there is the measurement error in the outcome of the parity measurement to determine the quadrature $\cos(\theta_{P, v})$ and $\sin(\theta_{P, v})$.

In order to ensure that the quantum state finally prepared between the site $v$ and the reference beam does not deviate significantly from the ideal state $\ket{\Phi_{P, v}}$ defined in Eq.~\ref{eq:phase_acc_NOON_ideal} due to the state errors in the first two stages of the protocol estimated in Eq.~\ref{eq:error_first_stage_calib} and \ref{eq:error_diffusion}, we choose $\varepsilon_0 \leq O(1 / P)$ and $T_\text{on} \leq O(1 / \sqrt{P})$. Note if we use the quantum enhanced twin-Fock state interferometry to determine the single-particle parameters, as described in section, the number of photons required in this stage will be $O(\varepsilon_0^{-1}) = O(P)$ which is only a constant-factor overhead. Assuming that we use $O(1)$ measurements of the parity operator to reduce its error to $\leq O(1)$, and accounting for the other sources of \ref{eq:error_first_stage_calib} and \ref{eq:error_diffusion}, $\theta_{v, P}$ can be determined to precision $\delta \theta_{P, v}$ given by
\[
\delta \theta_{P, v} \leq O(1) +O(\sqrt{P\varepsilon_0}) + O(\sqrt{P} T_\text{on}) \leq O(1),
\]
where we have suppressed the dependence on $N$ and $T_\text{prop}$, since they are independent of $P$. Thus, we can now estimate the error $\varepsilon$ in $ \chi_v$ as
\[
\varepsilon \sim \frac{\delta \theta_{P, v}}{d\theta_{P, v}/d\chi_v} \sim O(T_\text{on}^{-1}P^{-2}) \sim O(P^{-3/2}),
\]
which yields the super-Heisenberg scaling of $P = O(\varepsilon^{-2/3})$, improving over the $O(\varepsilon^{-2})$ of the classical tomography protocol for this problem.

\vspace{5pt}
\section{Conclusion}
 In this paper, we developed an algorithm for tomography of non-linear topological quantum bosonic lattices with measurement accessibility only to the perimeter of the lattice. We numerically demonstrated the stability of our algorithm and then proposed an extension to the algorithm to quantum mechanically enhance the precision of the estimated Hamiltonian. This algorithm can also  likely be extended to included other experimentally relevant quantum simulation setups like spin chains or quantum emitters coupled to such photonic lattices. Furthermore, while we only numerically studied the stability of the reconstruction algorithm, making progress in its rigorous mathematical analysis would also help understanding the practical utility and limitations of this algorithm.

 \begin{acknowledgements}
     We would like to thank Alexey Gorshkov for helpful discussions. AM acknowledges support from the National Science foundation in form of grants NSF-QII-TAQS-1936100, and NSF-1845009.
 \end{acknowledgements}

\nocite{*}
\bibliography{ref_biblo}

\appendix
\onecolumngrid

\section{Details of the tomography algorithm: detailed algorithm}\label{apxB}
\subsection{Extracting $M_{\alpha, v, u}^{(1)}, M_{\vec{\alpha}, v, u}^{(2)}$ from transmission and two-photon correlation functions}
In this appendix, we provide details of how the transmission and two-photon correlation functions can be used to obtain the data on $M_{\alpha, v, u}^{(1)}$ and $M_{\vec{\alpha}, v, u}^{(2)}$. Before proceeding further, we note a basic fact about complex functions --- suppose $f(\omega)$ is a complex valued function of $\omega$ which is of the form
\[
f(\omega) = \frac{1}{\sqrt{2\pi}}\sum_{j = 1}^n \bigg(\frac{F_j}{i(E_j - \omega) + \Gamma_j/2}\bigg),
\]
where $A_j$ are complex numbers, $E_j, \gamma_j$ are real numbers and $E_j \neq E_k$ for $j \neq k$. Then, it follows from the Filter diagonalization method of Ref.~\cite{wall1995extraction} that the function $f(\omega)$, if known for all $\omega \in \mathbb{R}$, uniquely determines $F_j, E_j, \Gamma_j$.

In both of the following subsection, we consider the setup shown in Fig.   --- the lattice is coupled to an input port, with time-domain annihilation operator $b_\tau$, through the vertex $u$ and to an output port, with time-domain annihilation operator $c_\tau$, through the vertex $v$.  The input port is then excited through an incident multi-tone laser pulse with frequencies $\omega_1, \omega_2 \dots \omega_M$, amplitude $B_0$, pulse width $\tau_w$ and pulse shape $\zeta(\tau)$ --- this can be described by the coherent state
\[
\ket{\psi_\text{in}} = \exp\bigg(-\frac{1}{2}{\int_{-\infty}^\infty \abs{f_{B_0, \tau_w}(\tau)}^2 d\tau\bigg)} \sum_{k = 0}^\infty \frac{1}{k!}\bigg(\int_{-\infty}^\infty f_{B_0, \tau_w}(\tau) b_\tau^\dagger d\tau \bigg)^k \ket{G},
\]
where
\[
f_{B_0, \tau_w}(\tau) = B_0 \zeta\bigg(\frac{\tau}{\tau_w}\bigg)\bigg(\sum_{i = 1}^M e^{-i\omega_i \tau}\bigg).
\]
For concreteness, we take $\zeta(\tau)$ to be a smooth function that falls off to 0 as $\abs{\tau}\to \infty$ faster than any polynomial in $\tau$ (i.e.~$\xi$ is a Schwartz function) and fix its normalization by setting $\zeta(0) = 1$. It is convenient to define $\ket{\phi_\text{in}^{(k)}}$ via
\[
\ket{\phi_\text{in}^{(k)}} = \frac{1}{B_0^k} \bigg(\int_{-\infty}^\infty f_{B_0, \tau_w}(\tau)b_\tau^\dagger d\tau\bigg)^k \ket{G}.
\]
We perform a homodyne detection on the output port --- the same laser pulse, with possibly a controllable phase $\varphi$, is interfered with the output port through a 50-50 beam splitter, and either a transmission or a two-photon correlation function measurement is performed at (one of the) beam-splitter output ports.

To extract information about the single and two particle spectra, we will consider the regime of continuous-wave excitation and weak driving. The continuous wave limit would correspond to $\tau_w \to \infty$ and the weak drive limit would correspond to $B_0 \to 0$ --- however, as was pointed out in Ref.~\cite{trivedi2019photon}, the order of these limits is important. For e.g., taking the limit of $B_0 \to 0$ before $\tau_w \to \infty$ simply results in $\ket{\psi_\text{in}}$ reducing to the vacuum state $\ket{G}$, while taking the limit of $\tau_w \to \infty$ without taking $B_0 \to 0$ does not exist. The resolution of this issue, as pointed out in Ref.~\cite{trivedi2019photon}, was to consider the transmission and the second order correlation functions of the scattered photons normalized to $B_0^2$ and $B_0^4$ respectively, then take the weak driving limit ($B_0 \to 0$) followed by the continuous wave limit $\tau_w \to \infty$.

Below, we execute this procedure for our setup and for first and second order correlation functions obtained in a Homodyne detection setup. In particular, we establish that the outcome of this detection allows us to obtain $E_\alpha^{(1)}, E_\alpha^{(2)}, M_{\alpha, v, u}^{(1)}, M_{\vec{\alpha}, v, u}^{(2)}$, where $v, u$ are vertices at the boundary. A subtlety that was implicitly assumed but not rigorously outlined in Ref.~\cite{trivedi2019photon} is that the effective Hamiltonian within the single and two-excitation subspaces, $\textbf{H}_\text{eff}^{(1)} \in \mathbb{C}^{\abs{\mathcal{V}}\times \abs{\mathcal{V}}}$ and $\textbf{H}_\text{eff}^{(2)}\in \mathbb{C}^{\abs{\mathcal{V}}(\abs{\mathcal{V}} - 1)/2\times \abs{\mathcal{V}}(\abs{\mathcal{V}} - 1)/2}$, have all eigenvalues with negative imaginary parts i.e.~they do not support any \emph{bound states}. Under this assumption together with assuming that the matrices $\textbf{H}_\text{eff}^{(1)}, \textbf{H}_\text{eff}^{(2)}$ are diagonalizable, we then obtain that
\begin{align}\label{eq:upper_bound_norm_heff}
    \norm*{\exp\big({-i\textbf{H}_\text{eff}^{(k)}t}\big)} \leq C_ke^{-\sigma_k t},
\end{align}
for some constants $C_k, \sigma_k > 0$. To see this, we can first invoke the diagonalizability of $\textbf{H}_\text{eff}^{(k)}$ to express it as $\textbf{H}_\text{eff}^{(k)} = \textbf{P}^{(k)} \text{diag}(\boldsymbol{\lambda}^{(k)})\textbf{P}^{(k)^{-1}}$. Then, we have that
\[
\norm*{\exp\big({-i\textbf{H}_\text{eff}^{(k)}t}\big)} = \norm*{\textbf{P}^{(k)} \text{diag}(e^{-i\boldsymbol{\lambda}^{(k)}t})\textbf{P}^{(k)^{-1}}} \leq \norm*{\textbf{P}^{(k)}}\norm*{\textbf{P}^{(k)^{-1}}}\norm*{\exp({-i\boldsymbol{\lambda}^{(k)}t})}.
\]
Noting that $\norm{\exp({-i\boldsymbol{\lambda}^{(k)}t})} \leq e^{-\sigma_k t}$, where $\sigma_k$ is the smallest magnitude imaginary part of the eigenvalues $\boldsymbol{\lambda}^{(k)}$, and setting $C_k = \norm*{\textbf{P}^{(k)}}\norm*{\textbf{P}^{(k)^{-1}}}$, we obtain the bound in Eq.~\ref{eq:upper_bound_norm_heff}.

\subsubsection{Analyzing the homodyned transmission measurement}
We first consider performing a transmission measurement at the output port following the beam splitter, while using only one laser frequency $\omega_L$ (i.e.~$f_{B_0, \tau_w}(\tau) = B_0 \zeta(\tau / \tau_w) e^{-i\omega_L \tau}$). The incident photon flux at position $\tau$ in the input port is given by $\bra{\psi_\text{in}} b_\tau^\dagger b_\tau \ket{\psi_\text{in}} = B_0^2 \abs{\zeta(\tau / \tau_w)}^2 \to B_0^2$ as $\tau_w \to \infty$. The transmission in the output waveguide in the weak-driving limit, on accounting for the homodyne measurement and normalizing to the input photon flux, is given by 
\begin{align*}
T(\omega_L) &=  \lim_{\tau_w\to \infty}\lim_{B_0 \to 0} \frac{1}{2B_0^2} \bra{\psi_\text{in}}\hat{\text{S}}^\dagger \bigg(c_\tau^\dagger + e^{-i\varphi}f_{B_0, \tau_w}^*(\tau)\bigg) \bigg(c_\tau + e^{i\varphi}f_{B_0, \tau_w}(\tau)\bigg) \hat{\text{S}} \ket{\psi_\text{in}}, \nonumber\\
&= \lim_{\tau_w \to \infty} \frac{1}{2}\abs*{\bra{G} c_\tau \hat{\text{S}}\ket{\phi_\text{in}^{(1)}} + \zeta\bigg(\frac{\tau}{\tau_w}\bigg)e^{i(\varphi-\omega_L \tau)} }^2, \nonumber\\
&=\frac{1}{2}\abs*{ \lim_{\tau_w \to \infty}\bra{G} c_\tau \hat{\text{S}}\ket{\phi_\text{in}^{(1)}} + e^{i(\varphi-\omega_L \tau)} }^2.
\end{align*}
where $\tau$ is any time-point --- as we will see below, in the continuous-wave limit, any dependence on $\tau$ of the transmission will vanish. An expression for $\bra{G} c_\tau \hat{\text{S}} \ket{\phi_\text{in}^{(1)}}$ can be obtained from standard scattering theory \cite{trivedi2018few, xu2015input} to obtain
\begin{align}\label{eq:def_integral_singleparticle}
\bra{G} c_\tau \hat{\text{S}} \ket{\phi_\text{in}^{(1)}} = -\sqrt{\gamma_c \gamma_b} \int_{-\infty}^\infty \bra{G}\mathcal{T}\big[\tilde{a}_v(\tau)\tilde{a}_u^\dagger(s)\big]\ket{G} \zeta\bigg(\frac{s}{\tau_w}\bigg) e^{-i\omega_L s}ds,
\end{align}
where $\tilde{a}_v(s) = e^{i\tilde{H}_\text{eff}s} a_v e^{-i\tilde{H}_\text{eff}s}$, where $\tilde{H}_\text{eff}$ also includes non-Hermtian terms arising due to the coupling of $a_v, a_u$ with the input and output ports i.e.~$\tilde{H}_\text{eff} = H_\text{eff} - i(\gamma_b a_u^\dagger a_u + \gamma_c a_v^\dagger a_v)/2$, where $H_\text{eff}$ is defined in Eq.~\ref{eq:effective_hamiltonian_introduction}. We now taking the continuous-wave limit of $\tau_w \to \infty$ --- this requires us to swap the order of the limit and integral in Eq.~\ref{eq:def_integral_singleparticle}. The legitimacy of this swap follows from the dominated convergence theorem --- to see this, we first note that 
\[
\bra{G} \mathcal{T}\big[\tilde{a}_v(\tau) \tilde{a}_u^\dagger(s) \big]\ket{G} = \begin{cases}
    \textbf{e}_v^\text{T}\exp\big({-i\textbf{H}_\text{eff}^{(1)}(\tau - s)}\big) \textbf{e}_u & \text{ if } s \leq \tau, \\
    0 & \text{ otherwise},
\end{cases}
\]
where $\textbf{e}_v \in \mathbb{C}^{\abs{\mathcal{V}}}$ is a unit vector with a $1$ at the location of the vertex $v$. Furthermore, it follows from Eq.~\ref{eq:upper_bound_norm_heff} that for all $\tau_w > 0$ and $s \leq \tau$,
\[
\abs*{\textbf{e}_v^\text{T}\exp\big({-i\textbf{H}_\text{eff}^{(1)}(\tau - s)}\big) \textbf{e}_u \zeta\bigg(\frac{s}{\tau_w}\bigg) e^{-i\omega_L s}} \leq \norm*{\exp\big({-i\textbf{H}_\text{eff}^{(1)}(\tau - s)}\big)} \abs*{\zeta\bigg(\frac{s}{\tau_w}\bigg)} \leq C_1 e^{-\sigma_1 (\tau - s)} \norm{\zeta}_\infty,
\]
where $\norm{\zeta}_\infty = \sup_{t \in \mathbb{R}} \abs{\zeta(t)}$. Thus, for all $\tau_w > 0$, the integrand in Eq.~\ref{eq:def_integral_singleparticle} is upper bounded by the function $g(s) = C_1 e^{-\sigma_1(\tau - s)} \norm{\zeta}_\infty$ if $s \leq \tau$ and $0$ if $s > \tau$, which is also absolutely integrable. Thus, the integral in Eq.~\ref{eq:def_integral_singleparticle} satisfies the conditions of the dominated convergence theorem, and consequently,
\begin{align*}
\lim_{\tau_w \to \infty}{\int_{-\infty}^\infty \bra{G}\mathcal{T}\big[\tilde{a}_v(\tau)\tilde{a}_u^\dagger(s)\big]\ket{G} \zeta\bigg(\frac{s}{\tau_w}\bigg) e^{-i\omega_L s} ds} &= \int_{-\infty}^\infty  \bigg(\lim_{\tau_w\to \infty}\bra{G}\mathcal{T}\big[\tilde{a}_v(\tau)\tilde{a}_u^\dagger(s)\big]\ket{G} \zeta\bigg(\frac{s}{\tau_w}\bigg) e^{-i\omega_L s}\bigg) ds, \nonumber\\
& = \int_{-\infty}^\infty \bra{G}\mathcal{T}\big[\tilde{a}_v(\tau)\tilde{a}_u^\dagger(s)\big]\ket{G} e^{-i\omega_L s} ds.
\end{align*}

\noindent Evaluating this integral, we then obtain that
\begin{align}\label{eq:single_particle_s_limit}
\lim_{\tau_w\to \infty} \bra{G} c_\tau \hat{\text{S}} \ket{\phi_\text{in}^{(1)}} = e^{-i\omega_L \tau}\uptau(\omega_L), \text{ where }\uptau(\omega_L) = i\sqrt{\gamma_c \gamma_b} \textbf{e}_v^\text{T}\big(\tilde{\textbf{H}}^{(1)}_\text{eff} - \omega_L \textbf{I}\big)^{-1} \textbf{e}_u = i\sqrt{\gamma_c \gamma_b} \sum_{\alpha} \frac{M_{\alpha, v, u}^{(1)}}{E_\alpha^{(1)} - \omega}.
\end{align}
The transmission $T(\omega_L)$, in the continuous-wave limit, can now be expressed as
\[
T(\omega_L) = \frac{1}{2}\big| e^{i\varphi} + \uptau(\omega_L)\big|^2.
\]
By varying the phase $\varphi$ in the homodyning scheme, we can $\uptau(\omega)$ as a function of $\omega$, which then allows us to extract the single-excitation eigen-energies $E_\alpha^{(1)}$ and the coefficients $M_{\alpha, v ,u}^{(1)}$ for $v, u \in \mathcal{V}_0$, using the filter diagonalization method.
\subsubsection{Analyzing the homodyned equal-time two-particle correlation function}
\emph{Two-photon correlation function}. Next, we consider the two-particle correlation function, $G^{(2)}(\tau_1, \tau_2; \omega_{1}, \omega_2)$, measured at the output port on driving the input port with a laser with two tones at $\omega_1$ and $\omega_2$. In the weak drive limit, we can then express $G^{(2)}(\tau_1, \tau_2; \omega_1, \omega_2)$ as
\begin{align}\label{eq:g2_def_limit}
&G^{(2)}(\tau_1, \tau_2; \omega_1, \omega_2) \nonumber\\
&= \lim_{\tau_w \to \infty} \lim_{B_0\to 0} \frac{1}{4 B_0^4}\bra{\psi_\text{in}} \hat{\text{S}}^\dagger \prod_{i \in\{1, 2\}}\big(c_{\tau_i}^\dagger + e^{-i\varphi} f_{B_0, \tau_w}^*(\tau_i)\big) \prod_{i \in \{1, 2\}} \big(c_{\tau_i} + e^{i\varphi} f_{B_0, \tau_w}(\tau_i)\big)^2 \hat{\text{S}}\ket{\psi_\text{in}}, \nonumber\\
& = \lim_{\tau_w\to \infty}\frac{1}{4}\abs*{\frac{1}{2}\bra{G}c_{\tau_1}c_{\tau_2} \hat{\text{S}}\ket{\phi_\text{in}^{(2)}} + \sum_{i \in\{1, 2\}}\frac{e^{i\varphi}}{B_0} f_{B_0, \tau_w}(\tau_i) \bra{G}c_{\tau_{{i}^c}} \hat{\text{S}}\ket{\phi_\text{in}^{(1)}} + \frac{e^{2i\varphi}}{B_0^2} f_{B_0, \tau_w}(\tau_1) f_{B_0, \tau_w}(\tau_2)}^2, \nonumber\\
& = \frac{1}{4}\abs*{\mathcal{G}^{(c)}(\tau_1, \tau_2; \omega_1,\omega_2) + \prod_{i \in \{1, 2\}}\bigg(\sum_{j \in \{1, 2\}} (\tau(\omega_j) + e^{i\varphi}) e^{-i\omega_j \tau_i}  \bigg)}^2,
\end{align} 
where, in the summation, ${i}^c = 1$ if $i = 2$ and ${i}^c = 2$ if $i = 1$ and 
\[
\mathcal{G}^{(c)}(\tau_1, \tau_2; \omega_1, \omega_2) = \frac{1}{2}\bra{G}c_{\tau_1} c_{\tau_2}\hat{\text{S}}\ket{\phi^{(2)}_\text{in}} - \bra{G}c_{\tau_1}\hat{\text{S}}\ket{\phi^{(1)}_\text{in}}\bra{G}c_{\tau_2}\hat{\text{S}}\ket{\phi^{(1)}_\text{in}}.
\]
We now focus on computing the term $\bra{G} c_{\tau_1}c_{\tau_2} \hat{\text{S}}\ket{\phi_\text{in}^{(2)}}$. Without loss of generality, we assume that $\tau_1 \geq \tau_2$, since $G^{(2)}(\tau_1, \tau_2; \omega_L)$ is invariant under a swap of the two time indices $\tau_1, \tau_2$. Using the explicit expression for $\ket{\phi^{(2)}_\text{in}}$, we obtain from standard scattering theory \cite{trivedi2018few, trivedi2019photon} that
\begin{align}
    \bra{G} c_{\tau_1}c_{\tau_2} \hat{\text{S}}\ket{\phi_\text{in}^{(2)}} = \gamma_c \gamma_b \int_{-\infty}^\infty \int_{\infty}^\infty \bra{G} \mathcal{T}\big[\tilde{a}_v(\tau_1)\tilde{a}_v(\tau_2) \tilde{a}_u^\dagger(s_1) \tilde{a}_u^\dagger(s_2)\big] \ket{G}\Omega_{\omega_1, \omega_2}(s_1, s_2) \zeta\bigg(\frac{s_1}{\tau_w}\bigg)\zeta\bigg(\frac{s_2}{\tau_w}\bigg)ds_1 ds_2,
\end{align}
where $\Omega_{\omega_1, \omega_2}(s_1,s_2) = (e^{-i\omega_1 s_1} + e^{-i\omega_2 s_1})(e^{-i\omega_1 s_2} + e^{-i\omega_2 s_2})$. Next, we take the continuous-wave limit ($\tau_w \to \infty$) --- following our previous calculation, we would like to swap the order of this limit and the integral with respect to $s_1, s_2$. In order to justify this swap, we use the dominated convergence theorem for which we need an upper bound on the magnitude of $f(s_1, s_2) = \bra{G}\mathcal{T}[\tilde{a}_v(\tau_1) \tilde{a}_v(\tau_2) \tilde{a}_u^\dagger(s_1) \tilde{a}_u^\dagger(s_2)]\ket{G} \Omega_{\omega_1, \omega_2}(s_1, s_2)$. We begin by first noting that $f(s_1, s_2) = f(s_2, s_1)$, so we can assume $s_1 \geq s_2$ and rewrite
\[
\bra{G} c_{\tau_1}c_{\tau_2} \hat{\text{S}}\ket{\phi_\text{in}^{(2)}} = 2\gamma_c \gamma_b \int_{-\infty}^\infty \int_{ -\infty}^{s_1} \bra{G} \mathcal{T}\big[\tilde{a}_v(\tau_1)\tilde{a}_v(\tau_2) \tilde{a}_u^\dagger(s_1) \tilde{a}_u^\dagger(s_2)\big] \ket{G}\Omega_{\omega_1, \omega_2}(s_1, s_2) \zeta\bigg(\frac{s_1}{\tau_w}\bigg)\zeta\bigg(\frac{s_2}{\tau_w}\bigg)ds_2 ds_1.
\]
Now, for $s_1\geq s_2$,
\[
\bra{G} \mathcal{T}\big[\tilde{a}_v(\tau_1)\tilde{a}_v(\tau_2) \tilde{a}_u^\dagger(s_1) \tilde{a}_u^\dagger(s_2)\big] \ket{G} = \begin{cases}
    \textbf{e}_v^\text{T}e^{-i\textbf{H}_\text{eff}^{(1)}(\tau_1 - \tau_2)}\textbf{A}_v^{(2)} e^{-i\textbf{H}^{(2)}_\text{eff}(\tau_2 - s_1)}\textbf{A}_v^{(2)^\dagger} e^{-i\textbf{H}^{(1)}_\text{eff}(s_1 - s_2)} \textbf{e}_u & \text{ if } s_2 \leq s_1 \leq \tau_2 \leq \tau_1, \\
    \textbf{e}_v^\text{T}e^{-i\textbf{H}_\text{eff}^{(1)}(\tau_1 - s_1)}\textbf{e}_u \textbf{e}_v^\text{T}e^{-i\textbf{H}_\text{eff}^{(2)}(\tau_2 - s_2)}\textbf{e}_u &\text{ if }s_2 \leq \tau_2 \leq s_1 \leq \tau_1, \\
    0 & \text{ otherwise}.
\end{cases}
\]
where $\textbf{A}_v^{(i)}$ is the matrix corresponding to the operator $a_v$ when applied to the $i^\text{th}$ excitation subspace. Consequently, we now have that for all $\tau_w \geq 0$,
\[
\abs*{ \bra{G} \mathcal{T}\big[\tilde{a}_v(\tau)\tilde{a}_v(\tau) \tilde{a}_u^\dagger(s_1) \tilde{a}_u^\dagger(s_2)\big] \ket{G}\Omega_{\omega_1, \omega_2}(s_1, s_2)\zeta\bigg(\frac{s_1}{\tau_w}\bigg)\zeta\bigg(\frac{s_2}{\tau_w}\bigg)} \leq g(s_1, s_2),
\]
where
\[
g(s_1, s_2) = \begin{cases}2\norm{\zeta}_\infty^2 \norm{\textbf{A}_v^{(2)}}^2 C_1 C_2 e^{-\sigma_2(\tau - s_1)}e^{-\sigma_1(s_1 - s_2)} & \text{ if } s_2 \leq s_1 \leq \tau_2 \leq \tau_1, \\
2\norm{\zeta}_\infty^2C_1^2 e^{-\sigma_1(\tau_1 - s_1)} e^{-\sigma_1(\tau_2 - s_2)} & \text{if } s_2 \leq \tau_2 \leq s_1 \leq \tau_1, \\
0 & \text{ otherwise}.
\end{cases}
\]
and it can easily be checked that $g(s_1, s_2)$ is absolutely integrable. Consequently, from the dominated convergence theorem, we obtain that
\begin{align*}
&\lim_{\tau_w \to \infty} \int_{-\infty}^\infty \int_{ -\infty}^{s_1} \bra{G} \mathcal{T}\big[\tilde{a}_v(\tau_1)\tilde{a}_v(\tau_2) \tilde{a}_u^\dagger(s_1) \tilde{a}_u^\dagger(s_2)\big] \ket{G}\Omega_{\omega_1, \omega_2}(s_1,s_2) \zeta\bigg(\frac{s_1}{\tau_w}\bigg)\zeta\bigg(\frac{s_2}{\tau_w}\bigg)ds_2 ds_1, \nonumber \\
&\qquad = \int_{-\infty}^\infty \int_{ -\infty}^{s_1} \lim_{\tau_w \to \infty}\bra{G} \mathcal{T}\big[\tilde{a}_v(\tau_1)\tilde{a}_v(\tau_2) \tilde{a}_u^\dagger(s_1) \tilde{a}_u^\dagger(s_2)\big] \ket{G}\Omega_{\omega_1, \omega_2}(s_1,s_2)\zeta\bigg(\frac{s_1}{\tau_w}\bigg)\zeta\bigg(\frac{s_2}{\tau_w}\bigg)ds_2 ds_1, \nonumber\\
&\qquad =  \int_{-\infty}^\infty \int_{ -\infty}^{s_1} \bra{G} \mathcal{T}\big[\tilde{a}_v(\tau_1) \tilde{a}_v(\tau_2) \tilde{a}_u^\dagger(s_1) \tilde{a}_u^\dagger(s_2)\big] \ket{G}\Omega_{\omega_1, \omega_2}(s_1,s_2) ds_1 ds_2.
\end{align*}
Now, we can explicitly evaluate this integral by considering different time orderings in the integrand. In particular, we have that
\begin{align*}
    \int_{-\infty}^\infty \int_{-\infty}^{s_1} \bra{G}\mathcal{T}\big[\tilde{a}_v(\tau_1) \tilde{a}_v(\tau_2) \tilde{a}_u^\dagger(s_1) \tilde{a}_u^\dagger(s_2)] e^{-i(\omega_1 s_1 + \omega_2 s_2)} ds_1 ds_2 = I_1(\tau_1, \tau_2; \omega_1, \omega_2) + I_2(\tau_1, \tau_2; \omega_1, \omega_2),
\end{align*}
where 
\begin{align}
% \label{apxB:eq005}
	&I_1(\tau_1, \tau_2; \omega_1, \omega_2) = \int^{\tau_1}_{s_1=\tau_2} \int^{\tau_2}_{s_2=-\infty} e^{-i(\omega_1 s_1+\omega_2 s_2)} \langle G|{a}_{v} e^{-iH_\text{eff}(\tau_1 - s_1)}{a}_u^\dag\ket{G}\bra{G}{a}_{v}e^{-iH_\text{eff}(\tau_2 -s_2)}  {a}_u^\dag| G\rangle ds_1ds_2, \nonumber \\
 &\qquad =-\sum_{k_1, k_2} \frac{\langle  G | a_v |r^{(1)}_{k_2} \rangle \langle l^{(1)}_{k_2}| a^\dag_u | G\rangle}{(E^{(1)}_{k_1} - \omega_1)}   \frac{\langle  G | a_v |r^{(1)}_{k_1} \rangle \langle l^{(1)}_{k_1}| a^\dag_u | G\rangle}{(E^{(1)}_{k_2} - \omega_2)} \bigg(1-e^{-i(E^{(1)}_{k_1}-\omega_1)(\tau_1 - \tau_2)}\bigg)e^{-i(\omega_1 \tau_1 + \omega_2 \tau_2)},
 \\
& I_2(\tau_1, \tau_2; \omega_1, \omega_2) = \int^{\tau_2}_{s_1=-\infty} \int^{s_1}_{s_2=-\infty} e^{-i(\omega_1 s_1+ \omega_2 s_2)} \bra{G}{a}_{v}e^{-iH_\text{eff}(\tau_1 - \tau_2)} {a}_{v}e^{-iH_\text{eff}(\tau_2 - s_1)}{a}_u^\dag e^{-iH_\text{eff}(s_1 - s_2)} {a}_u^\dag \ket{G} dt_1dt_2, \nonumber \\
 &\qquad =- \sum_{k_1,k_2,k_3} \frac{ \langle  G | a_v | r^{(1)}_{k_1} \rangle \langle l^{(1)}_{k_1} | a_v | r^{(2)}_{k_2} \rangle\langle l^{(2}_{k_2} | a^\dag_u | r^{(1)}_{k_3} \rangle \langle l^{(1)}_{k_3} | a^\dag_u | G\rangle e^{-i(E^{(1)}_{k_1} - \omega_1)(\tau_1 - \tau_2)}} {(E^{(2)}_{k_2} - \omega_1 - \omega_2)(E^{(1)}_{k_3} - \omega_2)}e^{-i(\omega_1 \tau_1 + \omega_2 \tau_2)}  .
\end{align}
Therefore, we finally obtain that
\begin{align}\label{eq:corr_gfunc}
&\mathcal{G}^{(c)}(\tau_1, \tau_2; \omega_1, \omega_2) = \gamma_c \gamma_b \sum_{i, j}e^{-i(\omega_i + \omega_j )\tau_2}\times \nonumber\\
&\qquad \qquad \bigg(\sum_{\alpha_1, \alpha_2} \frac{M_{\alpha_1, v, u}^{(1)} M_{\alpha_2, v, u}^{(1)} e^{-iE_{\alpha_1}^{(1)}(\tau_1 - \tau_2)}}{(E_{\alpha_1}^{(1)} - \omega_i)(E_{\alpha_2}^{(1)} - \omega_j)}  - \sum_{\alpha_1, \alpha_2, \alpha_3}\frac{M_{(\alpha_1, \alpha_2, \alpha_3), v, u}^{(2)}e^{-iE_{\alpha_1}^{(1)} (\tau_1 - \tau_2)}}{(E_{\alpha_2}^{(2)} - \omega_i - \omega_j)(E_{\alpha_3}^{(1)} - \omega_j)}\bigg)
\end{align}
Using Eqs.~\ref{eq:g2_def_limit} and \ref{eq:corr_gfunc}, we can then obtain the following compact expression for $G^{(2)}(\tau_1, \tau_2; \omega_1, \omega_2)$,
\begin{subequations}
    \begin{align}
        &G^{(2)}(\tau_1, \tau_2; \omega_1, \omega_2) = \bigg| e^{-i\Delta \tau_2} \bigg(\frac{1}{2}T^{(2)}(\tau; \omega_1, \omega_1) + e^{-i\omega_1 \tau} (\tau(\omega_1) + e^{i\varphi})^2\bigg) + \nonumber\\
        &\qquad \qquad \qquad \qquad \quad \ e^{i\Delta \tau_2}\bigg(\frac{1}{2}T^{(2)}(\tau; \omega_2, \omega_2) + e^{-i\omega_2 \tau}(\tau(\omega_2) + e^{i\varphi})^2\bigg) + \nonumber \\
        &\qquad \qquad \qquad \qquad \quad \ \bigg(T^{(2)}(\tau; \omega_1, \omega_2) + (e^{-i\omega_1 \tau} + e^{-i\omega_2 \tau})(\tau(\omega_1) + e^{i\varphi})(\tau(\omega_2) + e^{i\varphi})\bigg)\bigg|^2.
    \end{align}
Here $\Delta = \omega_1 - \omega_2$, $\tau = \tau_1 - \tau_2$ and
\begin{align}
    T^{(2)}(\tau; \omega_1, \omega_2) =\sum_{i \in \{1, 2\}} \sum_{\alpha_1, \alpha_2} \frac{M_{\alpha_1, v, u}^{(1)} M_{\alpha_2,v, u}^{(1)}e^{-iE_{\alpha_1}^{(1)}\tau}}{(E_{\alpha_1}^{(1)} - \omega_i)(E_{\alpha_2}^{(1)} - \omega_{i^c})} - \sum_{\alpha_1, \alpha_2, \alpha_3}\frac{M_{(\alpha_1, \alpha_2, \alpha_3), v, u}^{(2)}e^{-iE_{\alpha_1}^{(1)} \tau}}{(E_{\alpha_2}^{(2)} - \omega_i - \omega_{i^c})(E_{\alpha_3}^{(1)} - \omega_i)},
\end{align}
where, again, $i^c = 1$ if $i = 2$ and $2$ if $i = 1$.
\end{subequations}

\emph{Extracting $M_{\vec{\alpha}, v, u}^{(2)}, E_{\alpha}^{(2)}$}. Next, we show how a measurement of $G^{(2)}(\tau_1, \tau_2; \omega_1, \omega_2)$, as a function of $\tau_1, \tau_2, \omega_1, \omega_2$ can be used to determine $M_{\vec{\alpha}, v, u}^{(2)}$ and $E_{\alpha}^{(2)}$. Our strategy would be to show that we can extract $T^{(2)}(\tau; \omega_1, \omega_2) $ as a function of $\tau, \omega_1, \omega_2$, which can then be used to obtain both $M_{\vec{\alpha}, v, u}^{(2)}$ and $E_{\alpha}^{(2)}$. First, we fix $\tau = \tau_1 - \tau_2, \omega_1, \omega_2$, with $\omega_1 \neq \omega_2$, and vary $\tau_2$. If $\Delta = \omega_1 - \omega_2 \neq 0$, then $G^{(2)}(\tau_1, \tau_2; \omega_1, \omega_2)$, as a function of $\tau_2$, will be of the form $f(\tau_2) = \abs{C_0 + C_1 e^{-i\Delta \tau_2} + C_{-1}e^{i\Delta \tau_2}}^2$ where $C_0, C_1, C_2 \in \mathbb{C}$. From this function, we can determine $\abs{C_0}$ --- to see this, note that
\[
f(\tau_2) = \abs{C_0}^2 + \abs{C_1}^2 + \abs{C_2}^2 + 2\text{Re}\big((C_0 C_1^* + C_{-1}C_0^*)e^{i\Delta \tau_2}\big) + 2\text{Re}\big(C_{-1}C_1^* e^{2i\Delta \tau_2}\big).
\]
This allows for the determination of $A_0 = \abs{C_0}^2 + \abs{C_1}^2 + \abs{C_{-1}}^2$, $A_1 = C_0 C_1^* + C_{-1}C_0^*$ and $A_2 = C_{-1}C_1^*$. From the coefficients $A_0, A_1, A_2$, we can then obtain $\abs{C_0}$ by solving $\abs{A_1}^2 + \abs{A_1^2 - 4\abs{C_0}^2 A_2^*} = 2\abs{C_0}^2(A_0 - \abs{C_0}^2)$. Therefore, we can obtain $\abs{T^{(2)}(\tau; \omega_1, \omega_2) + (e^{-i\omega_1 \tau} + e^{-i\omega_2 \tau})(\tau(\omega_1) + e^{i\varphi})(\tau(\omega_2) + e^{i\varphi})}^2$. We can assume that $\tau(\omega)$ has already been determined by the homodyned transmission measurement. Consequently, by varying $\varphi$, we can now determine $T^{(2)}(\tau; \omega_1, \omega_2)$. Since $M^{(1)}_{\alpha, v, u}$ and $E_\alpha^{(1)}$ are already known, this measurement allows us to determine
\[
\sum_{\vec{\alpha}} \frac{M^{(2)}_{\vec{\alpha}, v, u}e^{-iE_{\alpha_1}^{(1)}\tau}}{(E_{\alpha_2}^{(2)} - \omega_1 - \omega_2)(E_{\alpha_1}^{(1)} - \omega_1)},
\]
as a function of $\tau, \omega_1$ and $\omega_2$, from which both $E_\alpha^{(2)}$ and $M_{\vec{\alpha}, v, u}^{(2)}$ can be extracted.

\section{Quantum enhanced tomography}\label{apxC}
\subsection{Relationship between $\phi_{v, u}$ and $M_{\alpha, v, u}^{(1)}$}
\begin{figure*}[t]
	\centering
	\includegraphics[width=0.4\linewidth]{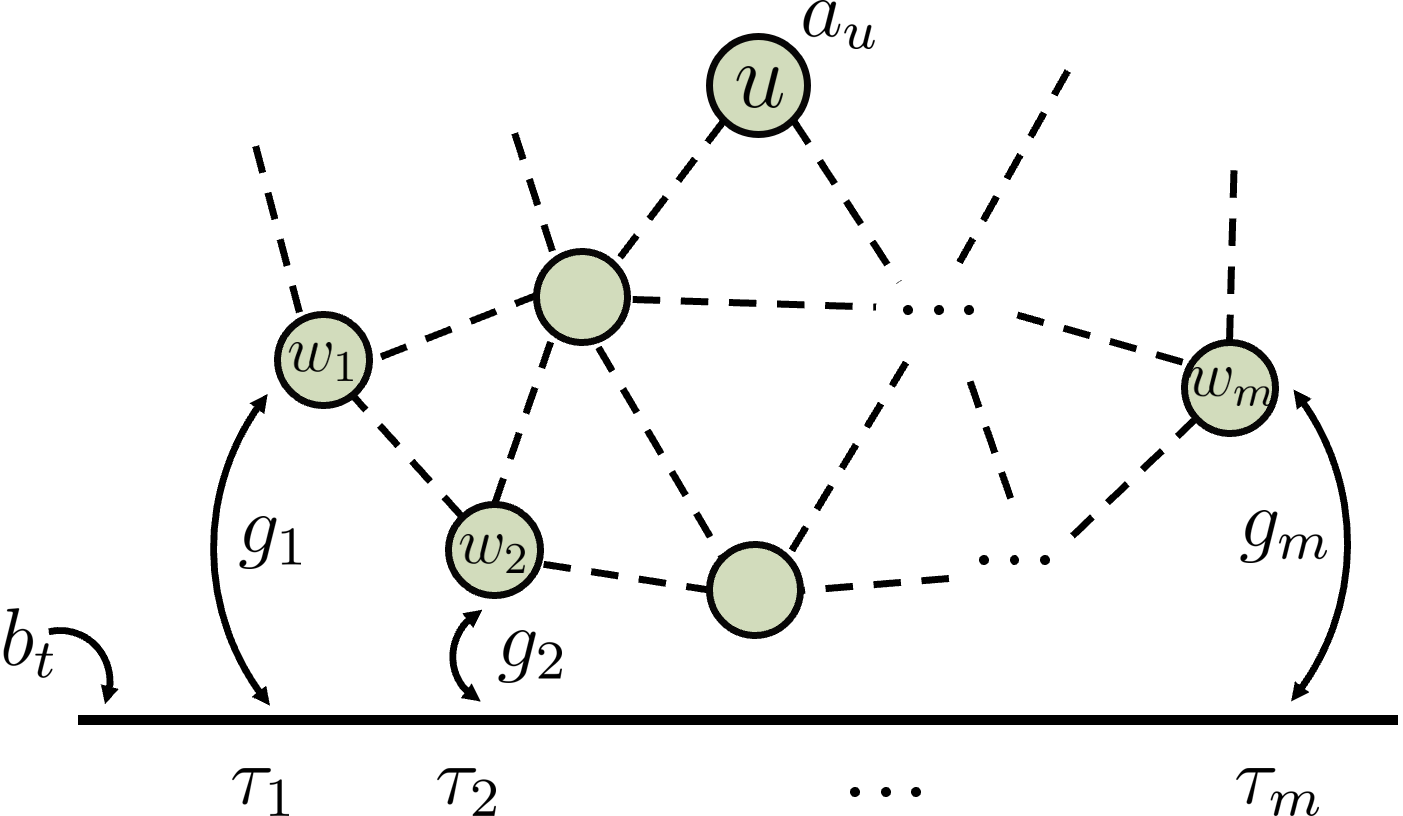}
	\caption{A schematic representation of the lattice, illustrating its coupling to the port through distinct sites denoted as $w_1, w_2, \ldots, w_m \in \mathcal{V}_0$. Each site $w_i$ is associated with a specific position along the port, characterized by the parameter $\tau_i$. The initial wave function within the port experiences scattering, driven by the Hamiltonian of both the port and lattice. When the nonlinearity is disabled, the output wave packet acquires a phase that depends on its frequency. \label{fig:appB1}}
\end{figure*}
Here, we consider the twin Fock-state MZI setup for measuring $\phi_{v, u}$ [Fig.~\ref{fig:enhancement_1}]. Note that the nonlinearity is switched off for this phase. We consider the lattice coupled to an output port through the sites $w_1,w_2,\cdots,w_m \in \mathcal{V}_0$ [Fig.~\ref{fig:appB1}]. The Hamiltonian of the system and the port in the interaction picture, relative to the Hamiltonian of the port, can be represented as:
\[
    H(t) = H_s + \bigg( \sum_{i=1}^{m} g_ia_{w_i} b^\dag_{t+\tau_i} +  \text{h.c.} \bigg),
\]
where:
\[
    H_s = \sum_{u,v=1}^N J_{u,v} a_u^\dag a_v.
\]
where, $g_i$ is the coupling rate of site $w_i$ to the port, and to simplify notation, we have assumed $J_{u,u} = \omega_u$. Here, $a_u$ is the annihilation operator of the bosonic mode at site $u$, and $b_t$ is the time domain annihilation operator corresponding to the coupling port. Furthermore, $\tau_i$ specifies the position of the port to which site $w_i$ is coupled and $m$ is the total number of sites coupled to the port. Writing the equations of motion for $b_\tau (t)$ and $a_u(t)$:
\begin{subequations}\label{eq:b_derivative}
\begin{align}
    &i\frac{db_\tau (t)}{dt} = \sum_{i=1}^m a_{w_i}(t) \delta(t+\tau_i-\tau) g_i, \\
    &i\frac{da_{w_i}(t)}{dt} = \sum_u J_{w_i,u} a_u(t) + b_{t+\tau_i} (t) g_i^*  \ \text{for }  i \in \{ 1,2,\cdots,m \},  \\
    &i\frac{da_v(t)}{dt} = \sum_u J_{v,u} a_u(t)  \ \text{otherwise}.
\end{align}
\end{subequations}
Integrating the first equation: 
\begin{align*}
    b_\tau (t) &= b_\tau (t_-) - i\sum_{i=1}^m g_i \int_{t_-}^t a_{w_i}(t') \delta(t'+\tau_i-\tau) dt', \\
    &= b_\tau(t_-) - i\sum_{i=1}^m g_i a_{w_i}(\tau -\tau_i) \Theta(t_- \leq \tau- \tau_i \leq t),
\end{align*}
where:
\[
    \Theta(a \leq t \leq b) = 
    \begin{cases}
        \frac{1}{2} \quad \text{if} \quad t \in \{a, b\}, \\ 
        1 \quad \text{if} \quad t \in (a, b), \\ 
        0 \quad \text{otherwise}.
    \end{cases}
\]
In particular,
\[
b_{t+\tau_j} (t) = b_{t+\tau_j}(t_-) - i\sum_{i=1}^m g_i a_{w_i}(t + \tau_j -\tau_i) \Theta(t_- \leq t + \tau_j - \tau_i \leq t).
\]
Assuming $\tau_1 < \tau_2 < \cdots < \tau_m$, we have that:
\begin{align} \label{eq:b_vs_a}
    b_{t+\tau_j} (t) = b_{t+\tau_j}(t_-) - i\frac{g_j}{2} a_{w_j} (t) - i \sum_{i=j+1}^{m} g_i a_{w_i}(t),
\end{align}
where we have taken $t_- \to -\infty$, and also used the fact that:
\begin{align*}
    \Theta(t_- \leq t + \tau_j - \tau_i \leq t) = 
    \begin{cases}
        1/2 \quad\text{if}\quad j=i, \\
        1 \quad\text{if}\quad j<i, \\
        0 \quad\text{otherwise}.
    \end{cases}
\end{align*}
and as a second step neglected delay $\tau_i \simeq \tau_j$. Using Equations \ref{eq:b_derivative}, and \ref{eq:b_vs_a},  we have that:
\[
    i\frac{da_{w_j}(t)}{dt} = \sum_u \big(\textbf{H}_\text{eff}\big)_{w_j,u} a_u(t) + g^*_j b_{t+\tau_j} (t_-) \quad\text{if}\quad j \in \{ 1,2,\cdots,m \},
\]
\[
    i\frac{da_v(t)}{dt} = \sum_u (\textbf{H}_\text{eff})_{u,v} a_u(t) \quad\text{otherwise},
\]
where:
\begin{align*}
    (\textbf{H}_\text{eff})_{w_j,w_j} &= J_{w_j,w_j} -i\frac{\abs{g_j}^2}{2} \quad \text{for}\quad j \in \{ 1,2,\cdots,m \}, \\
    (\textbf{H}_\text{eff})_{w_j,w_i} &= J_{w_j,w_i} -ig^*_jg_i \quad \text{for}\quad j,i \in \{ 1,2,\cdots,m \} \,\text{and}\, j<i, \\
    (\textbf{H}_\text{eff})_{u,v} &= J_{u,v} \quad \text{otherwise}. 
\end{align*}
We can integrate this equation explicitly: 
\begin{align}
    \begin{pmatrix}
        a_1(t) \\ a_2(t) \\ \vdots \\ a_N(t)
    \end{pmatrix} 
    = e^{-i\textbf{H}_\text{eff}(t-t_-)} 
    \begin{pmatrix}
        a_1(t_-) \\ a_2(t_-) \\ \vdots \\ a_N(t_-)
    \end{pmatrix}
    -i \sum_{n=1}^m \int_{t_-}^t e^{-i\textbf{H}_\text{eff}(t-s)} 
    g_n^* b_{s+\tau_n} (t_-) \textbf{e}_{w_n} ds.
\end{align}
where $\textbf{e}_{w_n}$ is a basis vector with a value of 1 at position $w_n$ and 0 elsewhere. Now, we are interested in:
\[
    S(\omega, \nu) =  [ b_\omega(t_+), b_\nu^\dag(t_-) ]  = \bra{0} b_\omega(t_+) b_\nu^\dag(t_-) \ket{0} = \frac{1}{2\pi}\int_{-\infty}^\infty\int_{-\infty}^\infty e^{i\omega\tau} e^{-i\nu \sigma} \bra{0} b_\tau (t_+) b^\dag_\sigma (t_-) \ket{0} d\tau d\sigma.
\]
Note that as $t_+ \to \infty$, and $t_- \to -\infty$, as well as neglecting the time delay by assuming $\tau_1 \simeq \tau_2 \simeq \cdots \simeq \tau_m \simeq = 0$, we can write:
\begin{align*}
    b_\tau (t_+) &= b_\tau(t_-) -i \sum_{i=1}^m g_i a_{w_i}(\tau-\tau_i) \Theta(t_- \leq \tau- \tau_i \leq t_+) \simeq b_\tau(t_-) -i \sum_{i=1}^m g_i a_{w_i}(\tau) \Theta(t_- \leq \tau \leq t_+), \\
    &= b_\tau (t_-) -i \left[ \textbf{g}^\text{T} e^{-i\textbf{H}_\text{eff}(\tau - t_-)}a(t_-) - i\sum_{n=1}^m \int_{t_-}^{\tau} \textbf{g}^\text{T} e^{-i\textbf{H}_\text{eff}(\tau - s)} 
    g_n^* b_{s+\tau_n} (t_-) \textbf{e}_{w_n}
    ds \right].
\end{align*}
where $\textbf{g}=\sum_{n=1}^m g_n \textbf{e}_{w_n}$. Therefore, 
\begin{align*}
     b_\tau (t_+) b_\sigma^\dag (t_-)  &= \delta(\tau - \sigma) - \sum_{n=1}^m \int_{t_-}^{\tau} \textbf{g}^\text{T} e^{-i\textbf{H}_\text{eff}(\tau - s)} g_n^* \textbf{e}_{w_n} \delta(s+\tau_n -\sigma) ds, \\ 
    &= \delta(\tau - \sigma) - \sum_{n=1}^m \textbf{g}^\text{T} e^{-i\textbf{H}_\text{eff}(\tau - \sigma +\tau_n)} g_n^* \textbf{e}_{w_n} \Theta(t_- \leq \sigma - \tau_n \leq \tau), \\
    &\simeq \delta(\tau - \sigma) - \textbf{g}^\text{T} e^{-i\textbf{H}_\text{eff}(\tau - \sigma)} \textbf{g}^* \Theta(t_- \leq \sigma \leq \tau).
\end{align*}
where we again assumed $\tau_n \simeq 0$. Inserting this into the scattering matrix, we obtain:
\begin{align*}
    S(\omega, \nu) &= \frac{1}{2\pi}\int_{-\infty}^\infty \int_{-\infty}^\infty \delta(\tau-\sigma) e^{i\omega\tau} e^{-i\nu\sigma} d\tau d\sigma - \frac{1}{2\pi} \int_{-\infty}^\infty\int_{-\infty}^\tau  \textbf{g}^\text{T} e^{-i\textbf{H}_\text{eff}(\tau - \sigma)} \textbf{g}^* e^{i\omega\tau} e^{-i\nu\sigma} d\tau d\sigma , \\ 
    &= \delta(\omega - \nu) - \frac{1}{2\pi}\int_{0}^\infty\int_{-\infty}^\infty \textbf{g}^\text{T} e^{-i\textbf{H}_\text{eff}t} \textbf{g}^* e^{i\omega t} e^{i(\omega-\nu)\sigma} dt d\sigma, \\ 
    &= \delta(\omega - \nu) \left[1- \int_{0}^\infty \textbf{g}^\text{T} e^{-i(\textbf{H}_\text{eff}-\omega I)t} \textbf{g}^* dt\right], \\ 
    &= \delta(\omega - \nu) \left[1 -i \textbf{g}^\text{T} \left(\textbf{H}_\text{eff}-\omega I\right)^{-1} \textbf{g}^*\right].
\end{align*}
This can be further expressed in terms of the eigenvectors and eigenvalues of the Hamiltonian to obtain:
\[
    S(\omega, \nu) = \delta(\omega - \nu) \bigg[1 +i \sum_{i,j, \alpha} \frac{g_ig_j^* M^{(1)}_{\alpha, w_i,w_j}}{\omega-E_\alpha} \bigg].
\]
\subsection{State Preparation for $\chi_v$ measurement}
The classical algorithm for obtaining the $\chi$ value was found to be unstable. In this section, we will demonstrate that if we have the capability to control the non-linearity, enabling us to turn it on and off at will, there exists a protocol that overcomes this instability. Suppose we want to measure the nonlinearity at a desired site $v \in \mathcal{V}$. First, we prepare the following NOON state in a two port wavegiude setup, 
\begin{align}\label{eq:NOON_init}
    \ket{\Psi_\text{I}} = \frac{1}{2} \bigg( \ket{G} \otimes \ket{P_{\text{out}}}_{\text{ref}} + e^{i\alpha} \ket{P_{\text{in}}} \otimes \ket{G}_\text{ref}\bigg),
\end{align}
where $\ket{G}$ is the zero photon state and $\alpha$ is a phase that we will pick later. Note that the refrence port is not coupled to the lattice. $\ket{P_{\text{out}}}$ ($\ket{P_{\text{in}}}$) is the P-photon Fock state resulting from initializing the system at $(a^\dag_v)^P\ket{G}/\sqrt{P!}$ and evolving it forward (back) in time. In appendix \ref{subsec:psi_vs_Pv}, we show how to compute the wave-packets $\psi_\text{in}$, and $\psi_\text{out}$ corresponding to the Fock states $\ket{P_\text{in}}$, and $\ket{P_\text{out}}$. In appendix \ref{subsec:gamma_vs_psi_in}, we further provide a concrete scheme to generate NOON states in \ref{eq:NOON_init}, given the wave-packets $\psi_\text{in}$, and $\psi_\text{out}$, starting from a NOON state in two harmonic oscillators and letting them controllably decay into the output ports. The next step is to let the wave function in Eq. \ref{eq:NOON_init} evolve. As expected the resulting wave function can be written as,
\[
    \ket{\Psi_{P,v}} = \frac{1}{2} \bigg( \ket{G} \otimes \ket{P_{\text{out}}}_{\text{ref}} + e^{i\alpha} \ket{P_{v}} \otimes \ket{G}_\text{ref} \bigg),
\]
where $\ket{P_v} = (a^\dag_v)^P\ket{G}/\sqrt{P!}$. Now if we turn on the nonlinearity for time $T_{\text{on}}$ and neglect the diffusion of photons to the other sites, $\ket{P_v}$ will simply acquire a phase, 
\[
    \ket{\Phi_{P, v}}=  \frac{1}{\sqrt{2}} \left(\ket{G}\otimes \ket{P}_\text{ref} + e^{i(\alpha - \theta_{v, P})}\frac{a_v^{\dag P}}{\sqrt{P!}}\otimes \ket{G}_\text{ref}\right),
\]
and, 
\[
    \theta_{P, v} = \bigg(\mu_v P + \frac{\chi_v P(P - 1)}{2}\bigg) T_\text{on},
\]
Finally, if let the system to decay into the port, the output wave function is, 
\[
    \ket{\Psi_\text{F}} = \frac{1}{2} \bigg( \ket{G} \otimes \ket{P_{\text{out}}}_{\text{ref}} + e^{i(\alpha - \theta_{v, P})} \ket{P_{\text{out}}} \otimes \ket{G}_\text{ref} \bigg).
\]
This state is then passed through a 50-50 Beam splitter, followed by a parity measurement on one of the Beam splitter's outputs. Assuming that the Beam splitter is described by $\begin{pmatrix}
    1/\sqrt{2} & 1/\sqrt{2} \\ 
    1/\sqrt{2} & -1/\sqrt{2}
\end{pmatrix}$, 
this measurement yields an expected value of $\cos{\theta_{v,P}}$ and $\sin{\theta_{v,P}}$ for the choice of $\alpha = 0, \pi/2$.

\subsubsection{Computing $\ket{\psi_{\text{in}}}$ to initialize the system at $\ket{P_v}$ }\label{subsec:psi_vs_Pv}
\begin{figure*}[t]
	\centering
	\includegraphics[width=0.7\linewidth]{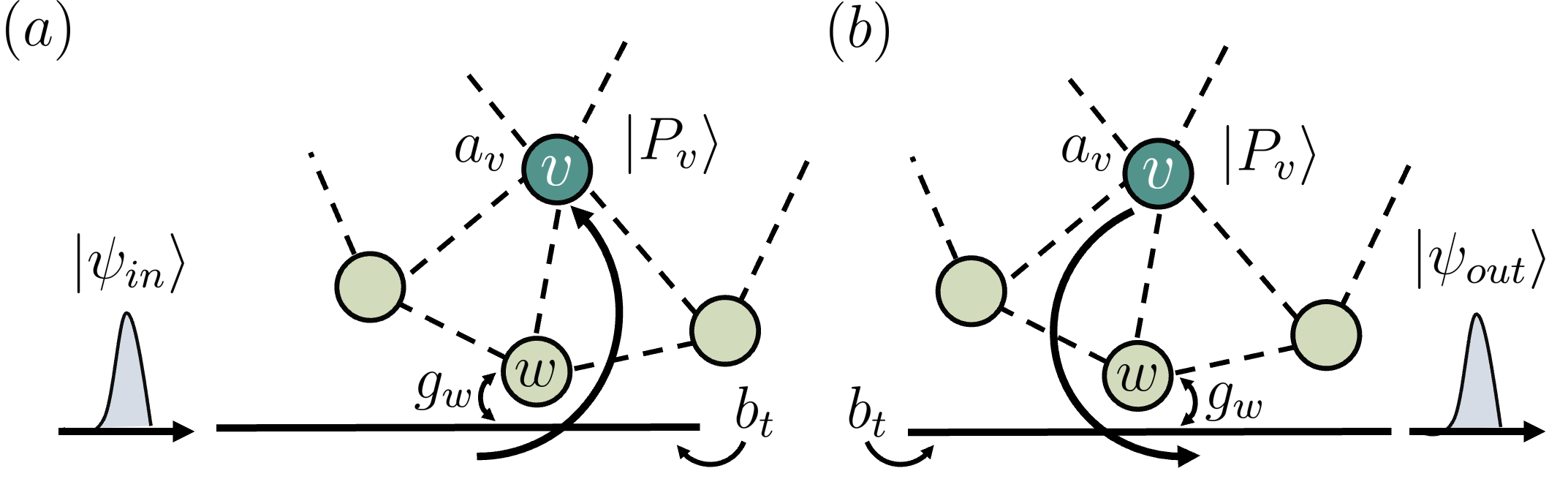}
	\caption{(a)  Initiating the lattice with a P-photon Fock state at site $v$ and tracing its evolution backward in time to determine the input wave packet within the port. (b) Illustrating the output wave packet, representing the state of the port following the initialization of the lattice with $\ket{P_v}$ and its forward time evolution. \label{fig:appB2a}}
\end{figure*}
Here, we begin by initializing the system with $P$ photons at site $v \in \mathcal{V}$, represented as $\ket{P_v}$. We then evolve the system backward (forward) in time to obtain the input (output) wave-packet [Fig.~\ref{fig:appB2a}]. The Hamiltonian governing the system and the port can be expressed as:
\[
    H = H_s + \int_{-\infty}^\infty \omega b^\dag_\omega b_\omega d \omega + \int_{-\infty}^\infty \big( A^\dag b_\omega + A b_\omega^\dag \big) \frac{d \omega}{\sqrt{2\pi}} \quad\text{and}\quad A = \sum_u g_u a_u,
\]
Here, $b_\omega$ is the annihilation operators for the port at frequency $\omega$ and $a_u$ is annihilation operator of the bosonic mode at site $u \in \mathcal{V}$. $g_u$ represents the coupling rate of site $u$ to the port, and it is only non-zero for sites that have a coupling connection to the port. The system's Hamiltonian, $H_s$, is given by:
\[
    H_s = \sum_{u} \omega_u a_u^\dag a_u + \sum_{u,u' \in \mathcal{E}} J_{u,u'}a_u^\dag a_{u'}.
\]
To analyze the system further, we work in the interaction picture with respect to the Hamiltonian of the port:
\[ 
    H_I = H_s + A^\dag b_t + Ab_t^\dag \quad\text{and}\quad b_t = \frac{1}{\sqrt{2\pi}} \int_{-\infty}^{\infty} b_\omega e^{-i\omega t} d \omega. 
\]
The Hamiltonian for both the system and port preserves the excitation number, and there is no photon-photon interaction since the nonlinearity is switched off. Therefore, we analyze the single particle problem with the initial state being $\ket{\psi(0)} = a_v^\dag \ket{0}$ for $v \in \mathcal{V}$. The evolved wave function at time $t$ can be represented as:
\begin{align}\label{eq:psi_t_for_1exc_multiple_couplings}
    \ket{\psi(t)} = \sum_u \alpha_u(t) a_u^\dag \ket{0} + \int_{-\infty}^{\infty} F_\tau (t) b_\tau^\dag \ket{0} d\tau,
\end{align}
with the initial condition values:
\[
    \boldsymbol{\alpha}(0) = \textbf{e}_v \quad\text{and}\quad F_\tau(0) = 0.
\]
Here, $\boldsymbol{\alpha}$ is the vector of $\alpha_u$, and $\textbf{e}_v$ is a basis vector with a value of 1 at position $v$ and 0 elsewhere. Our objective is to find the input wave packet, which is described by $F_\tau(-\infty)$. We can derive the equations of motion for $\boldsymbol{\alpha}$ and $F_\tau (t)$ as follows:
\begin{align} \label{eq:alpha_F_der}
    \frac{d\boldsymbol{\alpha}}{dt} = -i\textbf{J}\boldsymbol{\alpha} (t) - i\textbf{g}^* F_t(t), \\
    \frac{d F_\tau (t)}{d t} = -i\textbf{g}^\text{T}\boldsymbol{\alpha} (t) \delta (t-\tau).
\end{align}
Here, $\textbf{g}$ is a vector of $g_u$, and $\textbf{J}$ is a matrix with elements $\textbf{J}_{u,u'} = J_{u,u'}$. To simplify notation we have assumed $\textbf{J}_{u,u} = \omega_u$. Integrating the second equation, we obtain:
\begin{align*}
    F_\tau(t) &= F_\tau(0) -i \int_0^t \textbf{g}^\text{T}\boldsymbol{\alpha}(t') \delta (t'-\tau) dt', \\ 
    &= 0 + \frac{i}{2}\textbf{g}^\text{T}\boldsymbol{\alpha} (\tau) \Theta(t \leq \tau \leq 0),
\end{align*}
where:
\[
    \Theta(t \leq \tau \leq 0) = 
    \begin{cases}
        \frac{1}{2} \quad \text{if} \quad \tau \in \{t, 0\}, \\ 
        1 \quad \text{if} \quad \tau \in (t, 0), \\ 
        0 \quad \text{otherwise},
    \end{cases}
\]
Therefore:
\[
    F_t(t) = \frac{i}{2} \textbf{g}^\text{T}\boldsymbol{\alpha}(t).
\]
and:
\[
    \frac{d\boldsymbol{\alpha}(t)}{dt} = -i\textbf{J}\boldsymbol{\alpha}(t) + \frac{\textbf{g}^*\textbf{g}^\text{T}}{2} \boldsymbol{\alpha}(t) = -i \textbf{H}^\dag_\text{eff} \boldsymbol{\alpha}(t) \to \boldsymbol{\alpha}(t) = e^{-i\textbf{H}^\dag_\text{eff}t} \textbf{e}_v,
\]
where: 
\[
    \textbf{H}_\text{eff} = \textbf{J} - i \frac{\textbf{g}^*\textbf{g}^\text{T}}{2}.
\]
Thus:
\[
    F_\tau(t) = i \left( \textbf{g}^\text{T}e^{-i\textbf{H}^\dag_\text{eff} \tau}\textbf{e}_v \right) \Theta(t \leq \tau \leq 0),
\]
This results in:
\[
    F_\tau(-\infty) = i\textbf{g}^\text{T}e^{-i\textbf{H}^\dag_\text{eff} \tau}\textbf{e}_v,
\]
Similarly, we can determine the wavepacket for $t \to \infty$: 
\[
    F_\tau(\infty) = -i\textbf{g}^\text{T}e^{-i\textbf{H}_\text{eff} \tau}\textbf{e}_v.
\]
It's important to note that the Imaginary part of the eigenvalues of $\textbf{H}_\text{eff}$ are either zero or negative. Assuming there are no bound states in the system, there are no zero eigenvalues, indicating that $e^{-i\textbf{H}^\dag_\text{eff}\tau}$ ($e^{-i\textbf{H}_\text{eff}\tau}$) approaches zero as $\tau \to -\infty$ ($\tau \to \infty$). \\

Finally, for the case with $P$ photons, we can write the input and ouput wave function as: 
\[
    \ket{\psi_\text{in}} = \frac{1}{\sqrt{P!}} \bigg( \int_{-\infty}^\infty F_\tau(-\infty) b_\tau^\dag d\tau \bigg)^P \ket{0} 
    \text{ and }
    \ket{\psi_\text{out}} = \frac{1}{\sqrt{P!}} \bigg(\int_{-\infty}^\infty F_\tau(\infty) b_\tau^\dag d\tau \bigg)^P \ket{0}.
\]
\subsubsection{Relationship between $g(t)$ and $\ket{\psi_{\text{in}}}$} \label{subsec:gamma_vs_psi_in}
Consider a cavity that is coupled to a port, with the coupling coefficient $g(t)$ varying with time. Initially, the cavity is in an excited state and gradually emits a photon into the port. The objective is to determine the time-dependent parameter $g(t)$ in order to obtain the desired final wavefunction $\ket{\psi_{\text{in}}}$, expressed as:
\[
    \ket{\psi_{\text{in}}} = \int_{-\infty}^\infty F_\tau(\infty) b^\dag_\tau d\tau.
\]
In the interaction picture with respect to the port's Hamiltonian, the Hamiltonian can be represented as:
\[ 
    H(t) = g(t) b^\dag_t a + \text{h.c.} .  % + H_\text{c}
\]
Here, $a$ is the annihilation operator for the cavity mode, and $b_t$ is the annihilation operator for the port mode at time $t$. Consequently, we can describe the combined wavefunction of the cavity and the port as follows:
\[
    \ket{\psi(t)} = \alpha(t) a^\dag \ket{0} + \int_{-\infty}^\infty F_\tau(t) b^\dag_\tau \ket{0} d \tau .
\]
The initial conditions are set as:
\[
    \alpha(0) = 1 \quad\text{and} \quad F_\tau (0) = 0.
\]
The equation of motion is then given by:
\[
    \dot{\alpha} a^\dag \ket{0} + \int_{-\infty}^\infty \dot{F}_\tau(t) b^\dag_\tau \ket{0} d \tau = -i H \ket{\psi(t)}.
\]
By applying $\bra{0}a$ and $\bra{0}b_\tau$ from the left, we obtain:
\[
    \dot{\alpha} (t) = -i g^*(t) F_t(t) \text{ and }\dot{F}_\tau (t) = -i g(t) \alpha (t) \delta(t-\tau).
\]
Solving these equations results in 
\begin{equation*}
F_\tau (t) = 
    \begin{cases}
        0 & t < \tau,\\ 
        -\frac{i}{2}g(\tau) \alpha(\tau) & t = \tau, \\
        -ig(\tau) \alpha(\tau) & t > \tau,
    \end{cases} \text{ and }\alpha (t) = e^{-\frac{1}{2}\int_0^t \abs{g(s)}^2 d s}.
\end{equation*}
As $t$ approaches infinity, we can establish a relationship between $F_\tau(\infty)$ and $g(t)$ as follows:
\[
    F_\tau (\infty) = -ig(\tau) e^{-\frac{1}{2}\int_0^\tau  \abs{g(s)}^2 d s}.
\]
Finally, it can be demonstrated that:
\[
    g(\tau) = \frac{iF_{\tau}(\infty)}{\sqrt{1-\int_0^\tau \abs{F_s(\infty)}^2 ds}}.
\]
\subsection{Error analysis in $\chi_v$ measurement}
\subsubsection{Error in $\ket{P_v}$ due to linear tomography accuracy}
We can establish an upper bound on the error associated with initializing $P$ photons at any lattice site, taking into account deviations from the true values of the linear coefficients in the Hamiltonian. Consider the precision $\epsilon_0$ of the parameters $J_{u,v}$ and $\mu_{u}$ in the linear Hamiltonian $H^{(1)}$. We denote the measured linear Hamiltonian as $\hat{H}^{(1)}$. Suppose the input wavefunction is $\ket{\psi_{\text{in}}}$, which is described by the $F_\tau(-\infty)$ values derived in Section \ref{subsec:psi_vs_Pv}. We need to solve the equations \ref{eq:alpha_F_der}, with the initial condition values: 
\[
    \boldsymbol{\alpha}(-\infty) = 0 \quad\text{and} \quad \hat{F}_\tau(-\infty) = i\textbf{g}^\text{T}e^{-i\hat{\textbf{H}}^\dag_\text{eff} \tau} \textbf{e}_v,
\]
Therefore, 
\[
    F_t(t) = \hat{F}_t(-\infty) -i\int_{-\infty}^t \textbf{g}^\text{T} \boldsymbol{\alpha}(t') \delta(t'-t) dt' = \hat{F}_t(-\infty) -\frac{i}{2} \textbf{g}^\text{T}\boldsymbol{\alpha}(t).
\]
Inserting this into the derivative of $\boldsymbol{\alpha(t)}$:
\[
    \frac{d\boldsymbol{\alpha}}{dt} = -i\textbf{H}_\text{eff}\boldsymbol{\alpha} - i\textbf{g}^*\hat{F}_t(-\infty) = -i \hat{\textbf{H}}_\text{eff} \boldsymbol{\alpha} + i \left( \hat{\textbf{H}}_\text{eff} - \textbf{H}_\text{eff} \right) \boldsymbol{\alpha} - i\textbf{g}^*\hat{F}_t(-\infty),
\]
It should be noted that the true evolved wave function is obtained using the true Hamiltonian $\textbf{H}_\text{eff}$ instead of the reconstructed one. By solving the above equation:
\[
    \boldsymbol{\alpha}(t) = \int_{-\infty}^t e^{-i\hat{\textbf{H}}_\text{eff} (t-s)} \textbf{g}^* \hat{F}_s(-\infty) ds - \int_{-\infty}^t e^{-i\hat{\textbf{H}}_\text{eff} (t-s)} \left(\hat{\textbf{H}}_\text{eff} - \textbf{H}_\text{eff}\right) \boldsymbol{\alpha}(s) ds,
\]
at $t=0$:
\[
    \boldsymbol{\alpha}(0) - \textbf{e}_v = \int_{-\infty}^0 e^{i\hat{\textbf{H}}_\text{eff} s} \left(\textbf{H}_\text{eff} - \hat{\textbf{H}}_\text{eff}\right) \boldsymbol{\alpha}(s) ds = \int^{\infty}_0 e^{-i\hat{\textbf{H}}_\text{eff} s} \left(\textbf{H}_\text{eff} - \hat{\textbf{H}}_\text{eff}\right) \boldsymbol{\alpha}(-s) ds.
\]
It can be noticed that $\hat{\textbf{H}}\text{eff} - \textbf{H}\text{eff}$ is a sparse matrix with a maximum of $d$ nonzero elements in each row or column, where $d$ is the degree of the graph describing the system's couplings. The value for each of these elements is also bounded by the linear tomography error $\epsilon_0$. Therefore:
\[
    \norm{\hat{\textbf{H}}_\text{eff} - \textbf{H}_\text{eff}} \leq \left(d+1\right)\epsilon_0,
\]
Here, we introduce the additional assumption:
\[
    \norm{e^{-i\hat{\textbf{H}}_\text{eff} s}} \leq C_0 \lambda_0 e^{-\lambda_0 s}.
\]
with coefficients $C_0$ and $\lambda_0$, which can be a function of the total number of sites in the system $N$. This assumption implies that there are no bound states in the system, and any initial excitation in the system will decay exponentially to the port in time. Furthermore, $C_0 / \lambda_0$ can be used as the propagation time $T_\text{prop}$ for an excitation in the system to decay out of the lattice. Thus, considering $\norm{\boldsymbol{\alpha(s)}} \leq 1$:
\[
    \norm{\boldsymbol{\alpha}(0) - \textbf{e}_v} \leq \int_0^\infty \norm{e^{-i\hat{\textbf{H}}_\text{eff} s}} \norm{\hat{\textbf{H}}_\text{eff} - \textbf{H}_\text{eff}} ds \leq \frac{C_0}{\lambda_0} \epsilon_0 \left(d+1\right).
\]
Now, we are interested in the distance between:
\[
    \norm{\ket{P_v}-\ket{\hat{P}_v}} = \sqrt{2 \text{Re} \left(1-\bra{\hat{P}_v}P_v\rangle\right)},
\]
where, 
\[
    \ket{\hat{P}_v} = \frac{\left(a_v^\dag\right)^P}{\sqrt{P!}} \ket{0}, 
\]
\[
    \ket{P_v} = \frac{\left(\hat{a}_v^\dag\right)^P}{\sqrt{P!}} \ket{0} = \frac{\left(\boldsymbol{\alpha}^\text{T}(0)\textbf{a}^\dag + \int F_\tau(0) b_\tau^\dag d\tau\right)^P}{\sqrt{P!}} \ket{0},
\]
The second equation is written from Eq.\ref{eq:psi_t_for_1exc_multiple_couplings}. The inner product can be obtained as follows, 
\[
    \bra{P_v}\hat{P}_v\rangle = \frac{1}{P!} \bra{0} \left(a_v\right)^P \left(\hat{a}_v^\dag\right)^P \ket{0} = [a_v, \hat{a}^\dag_v ]^P = \left(\textbf{e}_v^\text{T} \boldsymbol{\alpha}(0)\right)^P,
\]
Here, we used Wick's Theorem. Intuitively, we should choose $P$ pairs between each $a_v$ and $\hat{a}_v$, and there are $P!$ ways to do that. We can write:
\[
    \bra{P_v}\hat{P}_v\rangle \geq 1-P\norm{\boldsymbol{\alpha}(0) - \textbf{e}_v},
\]
Inserting this in to the distance measure:
\begin{equation}\label{eq:error_prop}
    \norm{\ket{P_v}-\ket{\hat{P}_v}} \leq \epsilon_1 \quad\text{and}\quad \epsilon_1 = \sqrt{\frac{2C_0}{\lambda_0}\epsilon_0 \left(d+1\right) P}.
\end{equation}
This provides an upper bound on the error associated with initializing $P$ photons at any lattice site.
\subsubsection{Error in $\ket{\phi_{P,v}}$ due to photon diffusion to other sites}
After preparing the state $\ket{P_v}$, the non-linearity is activated for a duration of $T_{\text{on}}$. This time period should be sufficiently long to allow for the measurable accumulation of the phase induced by the non-linearity. However, it should also be short enough to prevent the $P$ photons from diffusing to neighboring sites. To quantify the latter condition, let $H$ be the full Hamiltonian that includes on-site anharmonicities, linear on-site terms, and coupling terms. We can define $H_\chi$ as the same Hamiltonian $H$ without the coupling terms, that is:\\
 \begin{equation*}
	H - H_{\chi} = \sum_{v,u \in \mathcal{E}} J_{v,u} a_{v}^\dag a_u.
\end{equation*}
The wavefunction at time $T_{\text{on}}$ evolving with the Hamiltonian $H_\chi$ can be expressed as,
\begin{equation*}
    \ket{\phi_{P, v}} =  e^{-iH_{\chi}T_{\text{on}}}  \ket{P_v} = e^{-i(\mu_v P + \chi_v \frac{P(P-1)}{2})T_{\text{on}}}  \ket{P_v}.
\end{equation*}
Since there are no coupling terms in $H_{\chi}$, all $P$ photons remain at site $u$ and the evolved wavefunction simply acquires a phase. When the system evolves under the full Hamiltonian $H$, the wavefunction at time $T_{\text{on}}$ is:
\begin{equation*}
    \ket{\hat{\phi}_{P, v}} =  e^{-iHT_{\text{on}}}  \ket{P_v}.
\end{equation*}
We can quantify the distance between $\ket{\hat{\phi}_{P, v}}$ and $\ket{\phi_{P, v}}$ as: 
\[
    \norm{\ket{\hat{\phi}_{P, v}}-\ket{\phi_{P, v}}}^2 = 2 \text{Re}\left(1-\bra{\hat{\phi}_{P, v}}{\phi_{P, v}}\rangle\right),
\]
using the derivative of $\bra{\hat{\phi}_{P, v}}{\phi_{P, v}}\rangle$ with respect to time, 
\[
    \frac{d}{ds}\bra{\hat{\phi}_{P, v}}{\phi_{P, v}}\rangle = \frac{d}{ds} \bra{P_v} e^{iH s}e^{-iH_\chi s}\ket{P_v}= i \bra{\hat{\phi}_{P, v}(s)} H - H_\chi\ket{\phi_{P, v}(s)}.
\]
Therefore, 
\[
    \norm{\ket{\hat{\phi}_{P, v}}-\ket{\phi_{P, v}}}^2 \leq 2 \int_0^{T_{\text{on}}} \abs{\bra{\hat{\phi}_{P, v}(s)} H - H_\chi\ket{\phi_{P, v}(s)}}  ds.
\]
We can express the term in the integral as:
\begin{equation*}
\begin{aligned}
	\norm{ \ket{\hat{\phi}_{P, v}} - \ket{\phi_{P, v}}}^2 &\leq 2 \int_0^{T_{\text{on}}}  \sum_{u,u'} \abs{J_{u',u}} (\, \abs{ \bra{\phi_{P, v} (s)}a_{u'}^\dag a_u\ket{\hat{\phi}_{P, v}(s)} }  ) \, d s, \\ 
    & = 2 \int_0^{T_{\text{on}}}  \sum_{u',u} \abs{J_{u',u}} (\, \abs{ \bra{P}a_{u'}^\dag a_u \ket{\hat{\phi}_{P, v}(s)} }   ) \, d s, \\ 
    & = 2 \int_0^{T_{\text{on}}}  \sum_{u \in \mathcal{N}_v } \abs{J_{v,u}} \sqrt{P} \, \abs{ \bra{P-1}a_u \ket{\hat{\phi}_{P, v}(s)} } \, d s, \\
    & \leq 2 \int_0^{T_{\text{on}}}  \sum_{u \in \mathcal{N}_u } J \sqrt{P} \sqrt{\bra{\hat{\phi}_{P, v}(s)}a_u^\dag a_{u} \ket{\hat{\phi}_{P, v}(s)}} \, d s .
\end{aligned}
\end{equation*}
For the first equality we use the fact that without the coupling terms in the Hamiltonian, the evolved wavefunction is just a phase applied on the state $\ket{P_v}$. $\bra{P_v}a_{u'}^\dag$ is zero unless $u' = v$, therefore in the second equality, we just keep the terms that are neighbors of $v$( $u \in \mathcal{N}_v$). For the last inequality we use the Cauchy–Schwarz inequality, and we consider $J$ to be the maximum value of $J_{u,v}$ over $u, v$. \\
\begin{align}
 \norm{ \ket{\phi_{P, v}} - \ket{\hat{\phi}_{P, v}}}^2 \leq   2 J \sqrt{P}  \int_0^{T_{\text{on}} } \sum_{u \in \mathcal{N}_v }  \sqrt{N_u(s)} \, d s.
    \label{dist2}
\end{align}
The number operator at site $u$ is defined as:
\begin{equation*}
    \hat{N}_u = a_u^\dag a_{u}  \quad\text{and}\quad N_u(s) = \bra{\hat{\phi}_{P, v}(s)}a_u^\dag a_{u} \ket{\hat{\phi}_{P, v} (s)}.
\end{equation*}
To find the upper bound on the number of photons at site $u$, 
we solve the differential equation for the operator $\hat{N}_u$ in the Heisenberg picture:  
\begin{equation*}
    \frac{d\hat{N}_u}{ds} = -i \left[\hat{N}_u, H \right] = i \sum_{v \in \mathcal{N}_u} \left(J_{u,v} a_u^\dag a_v + J_{v,u} a_v^\dag a_u\right). 
\end{equation*}
Therefore the upper bound for the derivative of number operator expectation is obtained as following: 
\begin{equation*}
\begin{aligned}
    \frac{d}{ds}{\bra{\hat{\phi}_{P, v}(s)} \hat{N}_u \ket{\hat{\phi}_{P, v}(s)}} \leq & \sum_{v \in \mathcal{N}_u} J \left(\abs{ \bra{\hat{\phi}_{P, v}(s)}a_u^\dag a_v \ket{\hat{\phi}_{P, v}(s)}} + \abs{ \bra{\hat{\phi}_{P, v}(s)}a_v^\dag a_u \ket{\hat{\phi}_{P, v}(s)}}\right), \\ 
    & \leq 2J \sum_{v \in \mathcal{N}_u} \sqrt{N_u(s) N_v(s)},
\end{aligned}
\end{equation*}
which implies
\begin{equation}
     \frac{d}{ds}{\sqrt{N_u(s)}} \leq 2J \sum_{v \in \mathcal{N}_u} \sqrt{N_v(s)}\implies \sqrt{N_u(s)}\leq 2J d \sqrt{P} s.
    \label{num}
\end{equation}
in which $d$ is the maximum number of couplings at each site. Inserting Eq. \ref{num} into Eq.\ref{dist2}, 
\begin{align}
    \norm{ \ket{\hat{\phi}_{P, v}} - \ket{\phi_{P, v}}}  \leq  \epsilon_2 \quad\text{and}\quad \epsilon_2 = \sqrt{2P}J d T_{\text{on}}.
    \label{eq:error_diffusion}
\end{align}
To ensure that the distance does not grow as a function of $P$, the evolution time should scale as $T_{\text{on}} \propto 1/{\sqrt{P}}$. This choice ensures that, for short durations of $T_{\text{on}}$, the diffusion of photons to other sites can be neglected.

\subsubsection{Parity measurement error}
Because of the error in the linear tomography of the Hamiltonian and the diffusion of the photons to the other neighbor sites, while the nonlinearity is on, the final wavefunction $\ket{\hat{\Psi}_\text{F}}$, will be different from the desired wave function $\ket{\Psi_\text{F}}$. The distance between these two can be computed using the equations \ref{eq:error_prop}, and \ref{eq:error_diffusion}, 
\begin{equation}\label{eq:error_total}
    \norm{\ket{\hat{\Psi}_\text{F}} - \ket{\Psi_\text{F}}} \leq \epsilon_3 \quad\text{and}\quad \epsilon_3 = 2\epsilon_1 + \epsilon_2.
\end{equation}
We define $\langle{O}\rangle_{\psi}$ to be the expectation value of the parity operator $O$ on the state $\ket{\Psi_\text{F}}$ which results in the sine or cosine of the phase $\theta_{P,v}$ based on choosing the value of $\alpha$. Now consider $O_k$ to be the average outcome of $k$ measurements of $O$ on the state $\ket{\hat{\Psi}_\text{F}}$.  With a probability of at least $1-\delta$, we can derive the following using Chebyshev's inequality,
\begin{equation}\label{eq:chebyshev}
    \text{Prob}\left(\abs{O_k-\langle{O}\rangle_{\hat{\psi}}} \leq \frac{1}{\sqrt{k \delta}}\right) \geq 1-\delta,
\end{equation}
where $\langle{O}\rangle_{\hat{\psi}}$ is the expectation value of the parity operator $O$ on the state $\ket{\hat{\Psi}_\text{F}}$. Using the triangle inequality and defining: 
\begin{equation*}
    \abs{O_k-\langle{O}\rangle_{\hat{\psi}}} \geq 
    \abs{O_k-\langle{O}\rangle_{\psi}} - \abs{\langle{O}\rangle_{\hat{\psi}} - \langle{O}\rangle_{\psi}}.
\end{equation*}
By using the result in Eq.~\ref{eq:error_total}, and $\norm{\hat{O}}=1$, 
\begin{align*}
    \abs{\langle{O}\rangle_{\hat{\psi}} - \langle{O}\rangle_{\psi}}  \leq 2 \norm{O} \norm{\ket{\hat{\Psi}_\text{F}} - \ket{\Psi_\text{F}}}  = 2\epsilon_3,
\end{align*}
Inserting this into Eq.~\ref{eq:chebyshev}:
\begin{align}\label{eq:meas_error}
    \text{Prob} \bigg( \abs{O_k-\langle{O}\rangle_{\psi}} \leq \epsilon_4 \bigg) \geq 1-\delta  \quad\text{and}\quad \epsilon_4 = 2\epsilon_3 + \frac{1}{\sqrt{k \delta}}.
\end{align}
Note that we interpret $O_k$ as the sine or cosine of $\left(\mu_v P + \frac{P(P-1)}{2} \chi_{v,\text{est}}\right)T_\text{on}$, with $\chi_{v, \text{est}}$ as the estimation of $\chi_v$.\\
Calculating the error for the $\chi_v$ value solely using sine or cosine functions is not straightforward. This difficulty arises because the derivative of the inverse cosine or sine can become infinite for specific values of $\chi_v$. To address this issue, we conduct $2k$ measurements and derive the following bounds based on Eq. \ref{eq:meas_error}:
\begin{align*}
        \abs{\cos{(\chi_1 \theta) - \cos{(\chi_2 \theta)} }} \leq \epsilon_4 \quad\text{and}\quad
        \abs{\sin{(\chi_1 \theta) - \sin{(\chi_2 \theta)} }} \leq \epsilon_4.
    \end{align*}
Here, $\chi_1$ represents the real value, and $\chi_2$ is the estimated value. If $\epsilon_4$ is small enough, then either $\abs{\cos{(\chi_2 \theta )}} \leq 1-\epsilon_4$ or $\abs{\sin{(\chi_2 \theta )}} \leq 1-\epsilon_4$. 
Consider the case where $\abs{\sin{(\chi_2 \theta )}} \leq 1-\epsilon_4$. We define $\varrho_i = \cos{(\chi_i \theta)}$. Note that for any $\rho \in [\rho_1, \rho_2]$
\begin{equation*}
    \abs{(1 - \varrho^2) - (1 - \varrho_2^2)} \leq 2 \abs{\varrho - \varrho_2} \leq 2\epsilon_4,
\end{equation*}
from which it follows that
\begin{align*}
 \abs{(1-\varrho^2)}  \geq \abs{(1-\varrho_2^2)} - 2\epsilon_4 \geq (1-\epsilon_4)^2 - 2\epsilon_3.
\end{align*}
Therefore,
\begin{equation*}
    \theta\abs{\chi_1 - \chi_2} = \int^{\varrho_2}_{\varrho_1} \frac{d \varrho}{\sqrt{1-\varrho^2}} \leq  \frac{\epsilon_4}{\sqrt{(1-\epsilon_4)^2 - 2\epsilon_4}},
\end{equation*}
Now, assuming $\epsilon_4$ is small enough (in particular $\epsilon_4 \leq \frac{1}{4}$), $\theta\abs{\chi_1 - \chi_2} \leq  4\epsilon_4$.
A similar analysis can be performed if $\abs{\cos{(\chi_2 \theta )}} \leq1- \epsilon_3$. Therefore with a probability of least $1-2\delta$, $\chi_{v, \text{est}}$ satisfies
\begin{align*}
\abs{\chi_{v,\text{est}} - \chi_v} \leq \Delta \quad\text{where}\quad \Delta = \frac{8\epsilon_4}{P(P-1)T_\text{on}} = \frac{8}{P(P-1) T_\text{on}}\bigg[ 4\sqrt{2\frac{C_0}{\lambda_0}PN(d+1)\epsilon_0} + 2\sqrt{2P}JdT_{\text{on}} + \frac{1}{\sqrt{\delta k}}\bigg].
\end{align*}
Now for $P\gg 1$, we choose $T_\text{on} \leq O(1/\sqrt{P})$,  $\epsilon_0 \leq O(1/P)$, $\delta, k = \Theta(1)$ and obtain that
\[
    P \simeq \bigg[\frac{16}{\Delta} \bigg(\frac{2}{T_\text{on}}\sqrt{\frac{2C_0}{\lambda_0}(d+1)\epsilon_0}+\sqrt{2}Jd\bigg)\bigg]^{2/3} \leq  O(\Delta^{-2/3}).
\]
\end{document}